\documentclass[a4paper]{aa}
\usepackage{times}
\usepackage{mathptmx}
\usepackage{fontenc}
\usepackage{graphicx}
\usepackage{multirow}
\usepackage{amssymb}
\usepackage{txfonts}
\usepackage{rotating}
\usepackage{natbib,twoopt}
\usepackage[bookmarks=false,colorlinks=true, citecolor=blue, backref=true,breaklinks=true]{hyperref} 
\bibpunct{(}{)}{;}{a}{}{,} 
\makeatletter
\newcommandtwoopt{\citeads}[3][][]{\href{http://adsabs.harvard.edu/abs/#3}%
{\def\hyper@linkstart##1##2{}%
\let\hyper@linkend\@empty\citealp[#1][#2]{#3}}}
\newcommandtwoopt{\citepads}[3][][]{\href{http://adsabs.harvard.edu/abs/#3}%
{\def\hyper@linkstart##1##2{}%
\let\hyper@linkend\@empty\citep[#1][#2]{#3}}}
\newcommandtwoopt{\citetads}[3][][]{\href{http://adsabs.harvard.edu/abs/#3}%
{\def\hyper@linkstart##1##2{}%
\let\hyper@linkend\@empty\citet[#1][#2]{#3}}}
\newcommandtwoopt{\citeyearads}[3][][]%
{\href{http://adsabs.harvard.edu/abs/#3}
{\def\hyper@linkstart##1##2{}%
\let\hyper@linkend\@empty\citeyear[#1][#2]{#3}}}
\makeatother

\begin{document}

\title{Exceptional AGN-driven turbulence inhibits star formation in the 3C~326N radio galaxy\thanks{Based on {\it Herschel} observations. {\it Herschel} is an ESA space observatory with science instruments provided by European-led Principal Investigator consortia and with important participation from NASA.}}
\author{P.~Guillard\inst{1,2,3,4}\thanks{email: guillard@iap.fr}, 
F.~Boulanger\inst{3}, 
M.~D.~Lehnert\inst{1}, 
G.~Pineau des For\^ets\inst{3,5},
F.~Combes\inst{5},
E.~Falgarone\inst{6},
J.~Bernard-Salas\inst{7}
}
\institute{
Institut d’Astrophysique de Paris, CNRS, UMR 7095, 98 bis Boulevard Arago, F-75014 Paris, France
\and
Sorbonne Universit\'es, UPMC Universit\'e Paris 06, 4 Place Jussieu, F-75005 Paris, France
\and
Institut d'Astrophysique Spatiale, UMR 8617, CNRS, Universit\'e Paris-Sud, Bat. 121, 91405 Orsay, France
\and
Spitzer Science Center, IPAC, California Institute of Technology, Pasadena, CA 92215, USA
\and
Observatoire de Paris, LERMA, UMR 8112, CNRS, 61 Avenue de l'Observatoire, 75014 Paris, France
\and
LERMA/LRA, Ecole Normale Supérieure and Observatoire de Paris, CNRS, 24, rue Lhomond, 75005 Paris, France
\and
Department of Physical Sciences, The Open University, Milton Keynes MK7 6AA, UK
}

\authorrunning{Guillard et al.}
\titlerunning{AGN-driven turbulence inhibits star formation in the 3C~326N radio galaxy}
\offprints{Pierre Guillard}

\date{Received  Feb 10, 2014 / Accepted Oct 10, 2014}

\abstract
{We detect bright [C{\sc ii}]$\lambda 158 \, \mu$m line emission from the radio
galaxy 3C~326N at $z=0.09$, which shows no sign of ongoing or
recent star formation ($\mbox{SFR} < 0.07$~M$_{\odot}$~yr$^{-1}$)
despite having strong H$_2$ line emission and a substantial
amount of molecular gas ($2\times 10^9$~M$_{\odot}$, inferred from
the modeling of the far-infrared (FIR) dust emission and the CO(1-0) line
emission).  The [C{\sc ii}] line is twice as strong as the 0-0S(1) 17$\mu$m
H$_2$ line, and both lines are much in excess of what is expected from UV
heating. We combine infrared \textit{Spitzer} and \textit{Herschel}
photometry and line spectroscopy with gas and dust modeling
to infer the physical conditions in the [C{\sc ii}]-emitting gas. The [C{\sc ii}] line, 
like rotational H$_2$ emission, traces a significant fraction (30 to 50\%) of the total molecular gas mass. This gas is warm ($70 < T < 100$~K) and at moderate densities $700 < n_{\rm H} < 3000$~cm$^{-3}$, constrained by both the observed [C{\sc ii}]-to-[O{\sc i}] and [C{\sc ii}]-to-FIR ratios. The [C{\sc ii}] line is broad, asymmetric, with a redshifted core 
component ($FWHM = 390$~km~s$^{-1}$) and a very broad
blueshifted wing ($FWHM = 810$~km~s$^{-1}$). 
The line profile of [C{\sc ii}] is similar to the profiles of the near-infrared H$_2$ lines and the Na~D optical absorption lines, and is likely to be shaped by a combination of rotation, outflowing gas, and turbulence. If the line wing is interpreted as an
outflow, the mass loss rate would be higher than $20\,$M$_{\odot}$~yr$^{-1}$, and
the depletion timescale close to the orbital timescale ($\approx 3\times 10^7$~yr). If true, we are observing this object at a very specific and brief time in its evolution, assuming that the disk is not replenished. Although there is evidence of an outflow in this source, we caution that the outflow rates may be overestimated because the stochastic injection
of turbulent energy on galactic scales can create short-lived, large velocity 
increments that contribute to the skewness of the line profile and mimic outflowing gas. The gas physical conditions raise the issue of the heating mechanism of the warm gas, and we show that the dissipation of turbulent energy is the main heating process.  
Cosmic rays can also contribute to the heating, but cannot be the dominant heating source because it requires an average gas density that is higher than the observational constraints. After subtracting the contribution of the disk rotation, we estimate the turbulent velocity dispersion of the molecular gas to be $120 < \sigma _{turb} < 330$~km~s$^{-1}$,  which corresponds to a turbulent heating rate that is higher than the gas cooling rate computed from the line emission. The dissipation timescale of the turbulent energy ($2\times10^7 - 10^8$~yrs)  is comparable to or larger than the jet lifetime or the dynamical timescale of the outflow, which means that turbulence can be sustained during the quiescent phases when the radio jet is shut off. The strong turbulent support maintains a very high gas scale height (0.3 to 4~kpc) in the disk. 
The cascade of turbulent energy can inhibit the formation of gravitationally bound structures on all scales, which offers a natural explanation for the lack of ongoing
star formation in 3C~326N, despite its having sufficient molecular gas
to form stars at a rate of a few solar mass per year. To conclude, the bright [C{\sc ii}] line 
indicates that strong AGN jet-driven turbulence may play a key role in enhancing the amount of molecular gas (positive feedback) but still can prevent star formation on galactic scales (negative feedback).
}


\maketitle

 \section{Introduction} 
\label{sec:introduction}

It has been hypothesized that supermassive black holes play an important role
in regulating and limiting their own growth and that of their host galaxy
\citep[e.g.,,][]{Silk1998}. This self-regulation cycle goes under the rubric
of ``AGN feedback''. This feedback has a rich theoretical legacy and is
often invoked to explain, for example, why massive early type
galaxies are ``old, red, and (mostly) dead'' and show apparently
``anti-hierarchical'' evolutionary behavior \citep[e.g.,][]{Thomas2005}. 
It has also been proposed as the origin of the correlation between 
black hole masses and kinematics and masses
of spheroids \citep[e.g.,][ and references therein]{Kormendy2013}.

However, the exact physical processes that underpin AGN feedback and
regulate or suppress the growth rates of black holes and massive galaxies
still have to be elucidated either observationally or theoretically. Of
several possibilities, the most energetically favored
are: energetic outflows with high entrainment rates driven by the
mechanical and radiative energy output of AGN through disk winds,
radio jets, and/or radiation pressure \citep[e.g.,][ and references
therein]{Fabian2012, Morganti2013, McNamara2014}; cosmic ray 
pressure and ionization both exciting and energizing the surrounding 
interstellar medium \citep[ISM; e.g.,][]{Sijacki2008, Jubelgas2008, Salem2013}; 
and/or AGN-generated turbulence \citep[e.g.,][]{Nesvadba2010, Sani2012, Guillard2012}. 
All of these processes are purported to prevent the collapse and cooling of ISM
and/or halo gas thus inhibiting efficient star formation and fueling 
the supermassive black hole.  Of course, other processes might also be 
important such as the difficulty of gas disks becoming self-gravitating
and the increased shear when they are embedded in spheroidal potentials
\citep{Martig2009, Martig2013}.

Models suggest that AGN feedback, primarily in low redshift massive galaxies, 
must suppress the cooling of halo gas, which 
starves the galaxies of cold, dense gas, inhibits both star formation and
efficient fueling of the supermassive black hole.  It is clear that AGN
provide sufficient energy, in principle, to do so \citep[e.g.,][]{Best2006,
Rafferty2008, Ma2013}.  However, a significant fraction of early-type
massive galaxies have a substantial amount of warm neutral and cold
molecular gas \citep{Young2011, Crocker2012, Serra2012, Bayet2013}, so simple
energy arguments may not be enough. To inhibit star formation and
efficient AGN fueling, many observational studies of AGN feedback
have focused on outflows \citep{Fischer2010, Feruglio2010, Dasyra2011,
Dasyra2012, Mullersanchez2011, Sturm2011, Combes2013, Labiano2013,
Alatalo2014, Morganti2013, Mahony2013, McNamara2014}.  Spectroscopic observations
of ionized, neutral, and molecular gas suggest that winds are
ubiquitous in galaxies hosting AGN \citep[e.g.,][]{Morganti2005,
Lehnert2011, Dasyra2011a, Alatalo2011, Guillard2012, Cicone2014,
ForsterSchreiber2014}. The mass outflow rates are such that the time
to remove the gas from the galaxies is relatively short, often less
than or approximately the same as an orbital dynamical time within the galaxy
\citep[e.g.,][]{Cicone2014}.

The generation of turbulence by the strong mechanical and radiative output
of AGN is probably an equally crucial aspect of suppressing star formation
\citep[e.g.,][]{Nesvadba2010, Sani2012, Guillard2012, Alatalo2014b}. 
A significant fraction of the
gas kinetic energy from the AGN feedback must cascade from bulk motions
down to the scales where it is dissipated.
This cascade can drive shocks into molecular clouds, enhancing molecular
line emission on galactic scales \citep{Guillard2009}.  Bright H$_2$ line
emission from shocks have been detected with the \textit{Spitzer Space
Telescope} for a number of radio galaxies \citep{Ogle2010, Guillard2012}.
The turbulence amplitude within molecular clouds of size of 50~pc, inferred from 
modeling the line emission is one order of magnitude larger than in the
Milky Way. Such a high turbulence has so far not been explored in current models
of star formation \citep{Krumholz2005, Federrath2012, Federrath2013}.

Perhaps the least well studied of any of the feedback processes is the
role of cosmic rays in regulating ISM pressure and the ionization and
chemistry of the dense gas in the immediate environments of AGN. The
energy released by accretion onto supermassive black holes may raise the cosmic
ray flux to much higher values than those observed in local star-forming 
galaxies such as the Milky Way.  The cosmic ray pressure is
also considered as a plausible cause of outflows on galactic scales
\citep{Salem2013, Hanasz2013}. Furthermore, cosmic rays may be the
main source of ionization and heating of the cold ISM in their host
galaxies. This is a plausible explanation of the high fraction of warm
molecular gas in some galaxies hosting AGN inferred from H$_2$ and CO
spectroscopy \citep{Ferland2008, Ogle2010, Mittal2011}. 
Some studies suggest that flaring of radio jets can enhance the output of cosmic rays 
in AGN \citep[e.g.,][]{deGasperin2012, Laing2013, Meli2013} and that $\approx 10\,$\% 
of the jet kinetic power may be converted into cosmic ray luminosity 
\citep{Gopal-Krishna2010, Biermann2012}. However, we
lack observations that quantify the interaction of cosmic rays with the
gas that depends on their propagation.

Disentangling all these plausible underlying physical mechanisms
related to AGN feedback is generally made difficult by the astrophysical
complexity of AGN hosts, where bright bolometric AGN radiation, complex
gas kinematics, and star formation almost always co-exist.  The radio-loud
galaxy 3C~326N at z$=$0.09 is a fortuitous exception as a gas-rich,
 massive early-type galaxy that is so strongly
dominated by the mechanical energy injected by the radio jet into the ISM
that gas heating from star formation and AGN bolometric radiation can
be neglected. The source contains a total molecular gas mass 
M(H$_2$) = 2$\times$10$^9$ \,M$_\odot$,
corresponding to a molecular gas surface density of $\rm 100\,$M$_\odot$~pc$^{-2}$
and yet has a star formation rate of $0.07\,$M$_\odot$~yr$^{-1}$ or
less \citep{Ogle2007}, which is $\sim$20 times less than the rate expected from
the Schmidt-Kennicutt law.  About half the molecular gas in 3C326~N
is warm ($\gtrsim$100 K) as seen in the shock-excited H$_2$ rotational
lines \citep{Ogle2007, Ogle2010}. The ratio between the mass of warm molecular gas (derived from mid-IR H$_2$ line observations) and the mass of H$_2$ derived from CO line emission (assuming a Galactic CO-to-H$_2$ conversion factor) is about ten times more in 3C~326\footnote{Similar mass ratios are found in the intergalactic medium of Stephan's Quintet \citep{Guillard2012}} than in star-forming
galaxies \citep{Roussel2007}. The kinetic luminosity of the gas and line emission
exceed the energy injection rates from star formation and AGN radiation,
leaving the mechanical energy and cosmic rays deposited by the radio source
 the most likely processes heating the gas \citep[][]{Nesvadba2010}.

The first steps in characterizing the complex kinematics of the
molecular gas in 3C~326N have been achieved by IRAM PdBI and VLT/SINFONI
ro-vibrational H$_2$ spectroscopy.  The CO(1-0) line detected with PdBI
is weak (1~Jy~km~s$^{-1}$) and marginally spatially resolved (with
a beam size of $2.5'' \times 2.1''$) \citep{Nesvadba2010}. The line
seems broad (FWHM = 350$\pm$100~km~s$^{-1}$), but unfortunately the low
signal-to-noise ratio of this detection makes it difficult to infer any
detailed information about the kinematics of the bulk molecular gas mass.
 The near-infrared (NIR) 1-0 S(3) H$_2$ line suggests the possible presence of
 a rotating gas disk of diameter $\sim$3~kpc.  Interestingly, the
 spatially resolved line profiles are generally broad and irregular 
 and cannot be explained solely by rotation \citep{Nesvadba2011a}.
 Moreover, the deep Na~D absorption with a broad blueshifted wing
 indicates an outflow of atomic gas with a terminal velocity of
 $\approx-$1800~km~s$^{-1}$.

All of this makes 3C~326N an excellent target for obtaining a relatively
detailed physical understanding of how mechanical AGN feedback affects the 
ISM of massive galaxies and ultimately the growth of these galaxies. An
important prerequisite for understanding how the energy injected by the
AGN into the ISM is being distributed among the gas phases is to have
direct observational constraints on each of these phases. 
In this paper we present the \textit{Herschel/PACS} spectroscopy of the 
[C{\sc ii}]$\lambda \,158\mu$m line, 
which is a very good tracer of the gas mass 
and the main coolant of the neutral medium not shielded from the ambient UV light, 
where the carbon is singly ionized by UV photons and cosmic rays.
 These observations are a unique way of constraining the kinematics of the bulk 
 of the gas mass, since so far most of the  information about the kinematics of 
 the molecular phase comes from near-IR H$_2$  line emission that traces 
 a very small fraction of the molecular gas mass (less than 1\%).

This paper is organized as follows. In Sect.~\ref{sec:observations} we present 
the photometric and spectroscopic data we use, and in Sect.~\ref{sec:kinematics} we 
describe the gas kinematics and compare the [C{\sc ii}] line profile to other tracers. 
In Sect.~\ref{sec:diagnostics}, we combine the observations with gas and dust 
modeling to infer which gas phase is probed by the [C{\sc ii}] line and to derive the dust 
and gas masses, as well as the 
physical conditions of the molecular gas in 3C~326N. In Sect.~\ref{sec:gasheating} we discuss 
 the main heating sources of this gas, and in Sect.~\ref{sec:outflow_turbulence} we interpret the 
 shape of the [C{\sc ii}] line profile. Section~\ref{sec:dynamical_state}  investigates what are the 
dynamical state and vertical scale height of the disk and whether they are compatible with the turbulent pressure support 
deduced from the observations. Section~\ref{sec:feedback} discusses the role that turbulence may play as 
a feedback mechanism, within the context of galaxy evolution. 
Finally in Sect.~\ref{sec:summary} we present a summary of our results. 

Throughout the paper we adopt a H$_0=$70 km s$^{-1}$, $\Omega_M=$0.3,
$\Omega_{\Lambda}=$0.7 cosmology. Unless specified, all the gas masses
quoted in this paper include Helium.

\section{Data reduction and observational results}
\label{sec:observations}

\subsection{[C{\sc ii}]$\lambda158\mu$m \textit{PACS} line spectroscopy}
\label{subsec:observations_spec}

\begin{figure}
\centering
\includegraphics[width=0.5\textwidth]{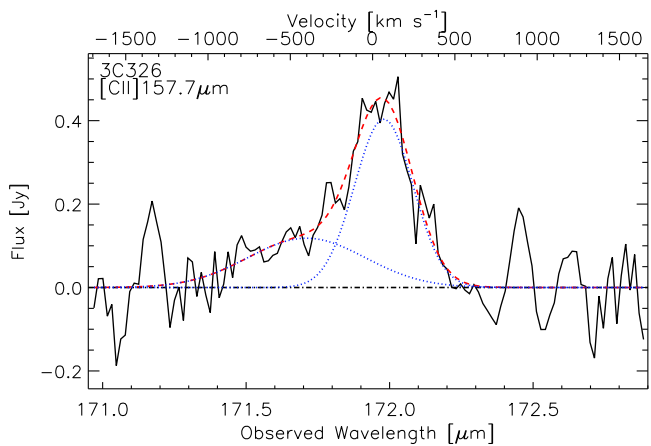}\\
\caption{[C{\sc ii}]$\lambda$158$\mu$m spectrum in 3C~326N observed with 
\textit{Herschel PACS} and obtained after combining the $3 \times 3$
central spaxels (see Sect.~2 for details). The reference recession velocity corresponds
to a redshift $z=0.09002$. The dotted blue lines show the two Gaussians
decomposition, and the red dashed line shows the  total fit. The fitted line parameters are gathered
in Table~\ref{tab:spectrum}, and the gas kinematics are discussed in
Sect.~\ref{sec:kinematics}. The two positive spurious features
on each side of the line (at $\lambda = 171.15 $ and $172.45  \,\mu$m)
have been masked to fit the line and the continuum.}
\label{fig:spectrum}
\end{figure}

\begin{table*}
\begin{center}
\begin{minipage}[t]{18cm}
 \renewcommand{\footnoterule}{}
\def\thefootnote{\alph{footnote}}
\caption{
[C{\sc ii}]$\lambda$158$\mu$m line measurements for 3C~326N. The
line is decomposed into two Gaussian components (a central component 
and blueshifted broad wing), see Figure~\ref{fig:spectrum}). Velocity
offsets are given relative to the optical redshift $z=0.09002$ and for a rest-frame wavelength
$\lambda_0$([C{\sc ii}])=157.741$\mu$m. The errors in the line
flux and luminosity include the statistical error from the fit and the
absolute flux calibration error of 12\% for the PACS R1 spectral band
as given in the PACS ICC Spectroscopy Performance and Calibration document.
The final error estimates are due to all uncertainties added in quadrature.}
\begin{tabular}{lcccccccc}
\hline \hline
line & wavelength\footnotemark[1] & shift\footnotemark[2]                   & peak flux\footnotemark[3]   & $\sigma$  & fwhm\footnotemark[4]               & fwhm-int \footnotemark[5]         & flux\footnotemark[6]  & luminosity \\
component     & [$\mu$m]     & [km s$^{-1}$]  & [mJy] & [$\mu$m] & [km s$^{-1}$] & [km s$^{-1}$] & [10$^{-17}$ W m$^{-2}$] & [$10^7$~L$_{\odot}$] \\
\hline 
central   & 171.98 & $67 \pm 4$   & $0.41 \pm 0.03$   &    $0.102\pm 0.003$   &  $418 \pm 12$  &  $343 \pm 10$   &    $0.74 \pm 0.10$  & $3.90 \pm 0.58$ \\
blue wing  & 171.71 & $-401 \pm 15$ & $0.12 \pm 0.01$   &    $0.204 \pm 0.005$   &  $839 \pm 21$  &  $804\pm 20$   &    $0.44 \pm 0.06$ & $2.29 \pm 0.34 $ \\ 
\hline 
total &          &      &       &       &       &       & $1.18\pm 0.10$ & $6.19 \pm 0.99 $ \\
\hline                                             
\end{tabular}
\footnotetext[1]{Central observed wavelength of the fitted Gaussian line.}
\footnotetext[2]{Velocity shift of the fitted Gaussian line with respect to the reference recession velocity of the galaxy.}
\footnotetext[3]{Peak line flux of the fitted Gaussian line.}
\footnotetext[4]{Full width at half maximum of the fitted Gaussian line.}
\footnotetext[5]{Intrinsic full width at half maximum of the fitted gaussian line, after subtraction (in quadrature) of the instrumental PACS resolution (239~km~s$^{-1}$ at $\lambda = 158\, \mu$m).}
\footnotetext[6]{Line integrated flux.}
\label{tab:spectrum}
\end{minipage}
\end{center}
\end{table*}

\begin{table*}
\centering
\caption{Far-infrared photometry for 3C~326~N. The fluxes quoted here are the measured fluxes (no color correction has been applied). }
\begin{tabular}{lccccc}
\hline \hline
instrument                    & \multicolumn{2}{c}{\textit{Spitzer / MIPS}} & \multicolumn{2}{c}{\textit{Herschel / PACS}} & \textit{Herschel / SPIRE} \\
wavelength [$\mu$m]  & 24   &  70       &  100   &    160   &    250  \\
bandwidth   [$\mu$m]  &  9   &   25      &    45   &     80    &    83 \\
\hline
flux [mJy]                      &  $0.52 \pm 0.07$ \footnotemark[1]    & $ 6.12 \pm 0.65$  & $16.7 \pm 0.6$ & $15.9 \pm 0.6$ & $9.0 \pm 3.5$ \\
\hline                                             
\end{tabular}
\footnotetext[1]{\citet{Dicken2011} reported a flux of $0.7 \ pm 0.1$~mJy.}
\label{tab:FIR_fluxes}
\end{table*}

We used the Photodetector Array Camera \& Spectrometer (PACS)
\citep{Poglitsch2010} onboard the \textit{Herschel Space Telescope} of ESA with
pointed observations in line spectroscopy mode to observe the redshifted
[C{\sc ii}]$\lambda \,158\, \mu$m line in 3C~326N at 172.22$\, \mu$m, corresponding to $z=0.09002$. Observations
were carried out on 24 July 2012.  We observed for a total of two cycles,
eight repetitions, and a total exposure time of 6079 seconds, including all
overheads. This allowed us to reach an rms sensitivity of 0.07~mJy.

We used the version 12.0 of the Herschel Interactive
Processing Environment \citep[HIPE][]{Ott2010} to reduce and calibrate
our data cube of 3C~326N. We also manually checked
 the flagging of bad pixels.  Building of the rebinned data cube
and averaging of nods were done using default parameters.  We also reduced
these data with a different pipeline that uses a different background
normalization method, the so-called flux-normalization method, which uses
the telescope background spectrum to normalize the signal \citep[see][for
a description of this approach]{Gonzalez-Alfonso2012}.  This allowed us
to check that the results were not affected by a possible uncertainty in
the relative spectral response function (RSRF) of the spectrometer. Both
methods gave very similar spectra (not more than 3\% relative difference)
and  identical fluxes within the absolute flux calibration rms uncertainty
(about 12\% in this wavelength range). We used the standard pipeline 
spectrum as our final product.

We extracted the spectrum within HIPE from the central spaxel by applying
the point-source flux correction recommended in the PACS manual. We
compared the results by summing the signals in the central $3\times3$
spaxels, and we found that this method gives a line flux 16\% higher
than the single central spaxel extraction.  We chose the $3\times3$
spaxels extraction for our final spectrum, because it minimizes the 
impact of pointing errors and the spatial extension of the source.
We also compared our results with the line fluxes given by the PACSman
software \citep{Lebouteiller2012}, and the results agreed
within the uncertainties.

We checked for a possible pointing error that could introduce
some artificial skewness in the line profile, but we find that the SDSS
R-band and 3mm continuum peaks are falling within less than 0.5 arcsec
of the center of the central spaxel. The offset is too small to be the
cause of the line asymmetry. We therefore conclude that it is real and
not an instrumental effect.

We note the presence of two prominent features at 171.15$\,\mu$m and 172.45$\,\mu$m. A reprocessing with the \textit{Herschel}/SAG4 pipeline developed at IAS did not make them disappear. They are likely to be artifacts or ripples and unlikely to be ghost lines because the wavelengths of those possible ghost lines, owing to the second pass in the optics of the PACS spectrometer, do not match any prominent line (for instance the [N{\sc ii}] line should have a ghost signal  at 178$\, \mu$m).

We fit the [C{\sc ii}] line with two Gaussians since the line
is strongly asymmetric, after masking the two spurious features at 171.15$\,\mu$m and 172.45$\,\mu$m (see Fig.~\ref{fig:spectrum}). The best-fit line
parameters are summarized in Table~\ref{tab:spectrum}.  We find a total
fine flux of $1.4 \pm 0.1 \times 10^{-17}$~W~m$^{-2}$, corresponding
to a [C{\sc ii}] line luminosity of  $7 \pm 1 \times 10^7$~L$_{\odot}$
for a luminosity distance of 410~Mpc. We defer the discussion of the
kinematics of the gas to Sect.~\ref{sec:kinematics}.

\subsection{Ancillary spectroscopic data}

We also used several already existing data sets to compare the neutral medium traced
by [C{\sc ii}] with other gas phases. This includes: imaging spectroscopy
of warm molecular Hydrogen obtained with VLT/SINFONI in the K-band,
which covers the Pa$\alpha$ as well as the H$_2$ 1$-$0 S(0) through S(5)
lines \citep{Nesvadba2011a}; \textit{Spitzer-IRS} mid-IR spectroscopy
\citep{Ogle2007, Ogle2010}, IRAM Plateau de Bure interferometry of CO(1-0)
and SDSS spectroscopy covering the Na~D line \citep{Nesvadba2010}. The main 
line measurements used in this paper are gathered in Table~\ref{tab:line_fluxes}.

The spectral resolving power of PACS at 170$\mu$m is 212~km~s$^{-1}$
(see PACS manual), about a factor~2 lower than that of SINFONI at 2.12
$\mu$m. To compare the line profiles at approximately the same spectral
resolution, we convolved the SINFONI line-spread function by a Gaussian
of $\sigma = 212$~km~s$^{-1}$.

\begin{table*}
\begin{minipage}[t]{\textwidth}
\caption{Summary of the 3C~326N emission line flux measurements used in this paper.}
\label{tab:line_fluxes}
\centering
\renewcommand{\footnoterule}{}  
\begin{tabular}{cccccc}
\hline \hline
\multirow{2}{*}{Instrument}   & \multirow{2}{*}{line}  &  rest wavelength & flux & FWHM & \multirow{2}{*}{reference} \\
  &   &  [$\mu$m]  &  [$10^{-18}$ W m$^{-2}$]  & [km s$^{-1}$] & \\
\hline
IRAM PdBI         & CO(1-0)              & 2600 & $0.0035 \pm 0.0007$ & $350 \pm 100$  & \citet{Nesvadba2011a}  \\
\textit{Herschel} PACS  & [CII]                      & 157.74     &  $11.8 \pm 1.0$       &   $372 \pm 12 $ \footnote{Since the line is highly asymmetric, we give a direct measurement of the line width at 50\% of the maximum flux.}      & This paper  \\
\textit{Herschel} PACS  & [OI]                      & 63.18        &  $6.1 \pm 1.3$       &  $352 \pm 31$       &  \citet{Guillard2014a} \\
\textit{Spitzer} IRS \footnote{For more rotational H$_2$ line and PAH measurements, see \citet{Ogle2010}.}         & H$_2$ 0-0 S(1)  & 17.1 &  $6.9 \pm 0.6$       &   Unresolved   &  \citet{Ogle2010}   \\
\textit{Spitzer} IRS         & PAH\footnote{Polycyclic aromatic hydrocarbons}  & 7.7 &  $ < 6 $       &   Undetected   &  \citet{Ogle2010}   \\
VLT SINFONI \footnote{For more ro-vibrational H$_2$ line and other near-infrared line measurements, see \citet{Nesvadba2011a}.}     & H$_2$ 1-0 S(3)  & 1.958 &  $2.2 \pm 0.2$      &  $594 \pm 41$    &  \citet{Nesvadba2011a} \\
VLT SINFONI     & Pa$\alpha$        & 1.875 & $1.20 \pm 0.14$       & $516 \pm 44$      &  \citet{Nesvadba2011a}  \\
SDSS \footnote{For more optical line measurements, see \citet{Nesvadba2010}.}                  & H$\alpha$          & 0.65646 &   $ 9.3 \pm 0.5$      &   $600 \pm 30$      & \citet{Nesvadba2010} \\
SDSS                 & H$\beta$           & 0.48627 &   $ 2.4 \pm 0.2$      &   $610 \pm 30$      & \citet{Nesvadba2010} \\
\hline                                             
\end{tabular}
\end{minipage}
\end{table*}

\subsection{\textit{PACS} and SPIRE infrared photometry}
\label{subsec:observations_phot}

The radio source 3C~326N has been observed with PACS at 100 and 160$\mu$m and with SPIRE
at 250$\mu$m, 350$\mu$m, and 500$\mu$m, as part of OT1\_pogle01 program 
(see Lanz et al., in prep., for the photometric measurements of the full sample of 
H$_2$-bright radio galaxies). 
We reprocessed the data from level 0 to level 2 maps with HIPE 10.0, 
using the standard naive map making with destripper.  
The source is detected at high signal-to-noise
ratios (10$\sigma$) on the PACS maps. We performed PACS photometry with
the SUSSEXtractor source extraction algorithm \citep{Savage2007}.
We find fluxes of $I_{\nu,\,100\mu \rm m} = 17.0 \pm 0.4 \,$mJy
and $I_{\nu,\,160\mu \rm m} = 15.6 \pm 0.4 \,$mJy. We compared our
results with a PSF (point spread function) fitting technique, using the FastPhot IDL library
\citep{Bethermin2010}, the PSF kernels computed by \citet{Aniano2012}
(after rebinning them and normalizing their area to 1), and the error
map given by the PACS pipeline. The results are compatible with the
source extraction photometry: we found $I_{\nu,\,100\mu \rm m} = 16.7
\pm 0.6 \,$mJy and $I_{\nu,\,160\mu \rm m} = 15.9 \pm 0.6 \,$mJy. We
adopt those fluxes for PACS.

In the SPIRE images, there is only a faint detection at 250$\mu$m.
We chose to perform the photometry by fitting a Gaussian to the
baseline-subtracted timeline samples on the sky, before the map-making
process, at the position of the source detected on the PACS 100$\mu$m
map.  We measured a point-source flux of $I_{\nu,\,250\mu \rm m} =
9.3 \pm 3.5 \,$mJy. For comparison, the PSF fitting photometry with
FastPhot gives $I_{\nu,\,250\mu \rm m} = 11.1 \pm 4.2 \,$mJy.  We also
performed a standard annular aperture photometry on the SPIRE map and
found $I_{\nu,\,250\mu \rm m} = 15.2 \pm 4.3 \,$mJy (after multiplying
by an aperture correction factor of 1.275). The photometry is limited
by the robustness of the background subtraction near the confusion limit
from IR galaxies, and aperture photometry on SPIRE maps gives poor
results in general.  Because of large error bars, the three measurements
are compatible, but given the PACS fluxes, we favor the result from the
timeline-fitting method, and we adopt a 250$\mu$m flux of 9~mJy. The
photometric results are summarized in Table~\ref{tab:FIR_fluxes}.
We use the far-infrared (FIR) fluxes to constrain the radiation field and
total gas mass (\S~\ref{subsec:dustmodel}).

\section{Kinematics of the gas} 
\label{sec:kinematics}

In this section, we describe the kinematics of the multiphase gas in
3C~326N, based on the [C{\sc ii}] observations and the comparison to
other data.

\subsection{Asymmetry of the [C{\sc ii}] line profile}
\label{subsec:kinematics_CII}

The [C{\sc ii}] line of 3C~326N appears asymmetric with a
pronounced blue wing (Fig.~\ref{fig:spectrum}).  The line widths
at 50\% and 20\% of the peak value are $\Delta \rm v _{50} = 372
\pm 9$~km~s$^{-1}$ and $\Delta \rm v _{20} = 769 \pm 18$~km~s$^{-1}$
respectively.  The line can be fit with two Gaussian profiles, one central component 
, slightly redshifted by $+$67~km s$^{-1}$, and one blueshifted by $-$400~km
s$^{-1}$ relative to the optical redshift at $z=0.09002$ (see Fig.~\ref{fig:spectrum}
and Table~\ref{tab:spectrum}).

The shape of the lines in 3C~326N is very likely a result of a
mixture of disk rotation, outflowing gas, and large-scale turbulence.
Inferring the relative contributions of these three components to
the [C{\sc ii}] line profile is not trivial, because the angular
resolution of our PACS data is insufficient for spatially resolving the
[C{\sc ii}] line, so we cannot directly determine where the various
possible components in the [C{\sc ii}] originate.  We must therefore rely
on the comparison of the integrated [C{\sc ii}] line profile to
other tracers (some being spatially resolved) to infer the astrophysical
processes shaping the [C{\sc ii}] line.

\subsection{Comparison of the line kinematics}
\label{subsec:kinematics_comparison}

\begin{figure}
\centering
\includegraphics[width=\columnwidth]{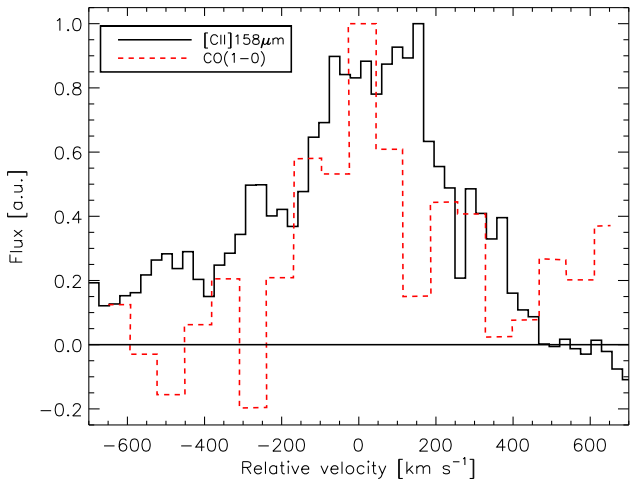}
\caption{Comparison of the [C{\sc ii}]$\lambda$158$\,\mu$m (solid black line) and CO(1-0)
(red dashed line) line profiles in 3C~326N. The maximum flux for both spectra has been normalized to 1 to ease the comparison. The CO(1-0) data are from IRAM Plateau de Bure observations reported in \citet{Nesvadba2010}. The reference recession velocity corresponds
to a redshift $z=0.09002$.}
\label{fig:c2co}
\end{figure}

\begin{figure}
\centering
\includegraphics[width=\columnwidth]{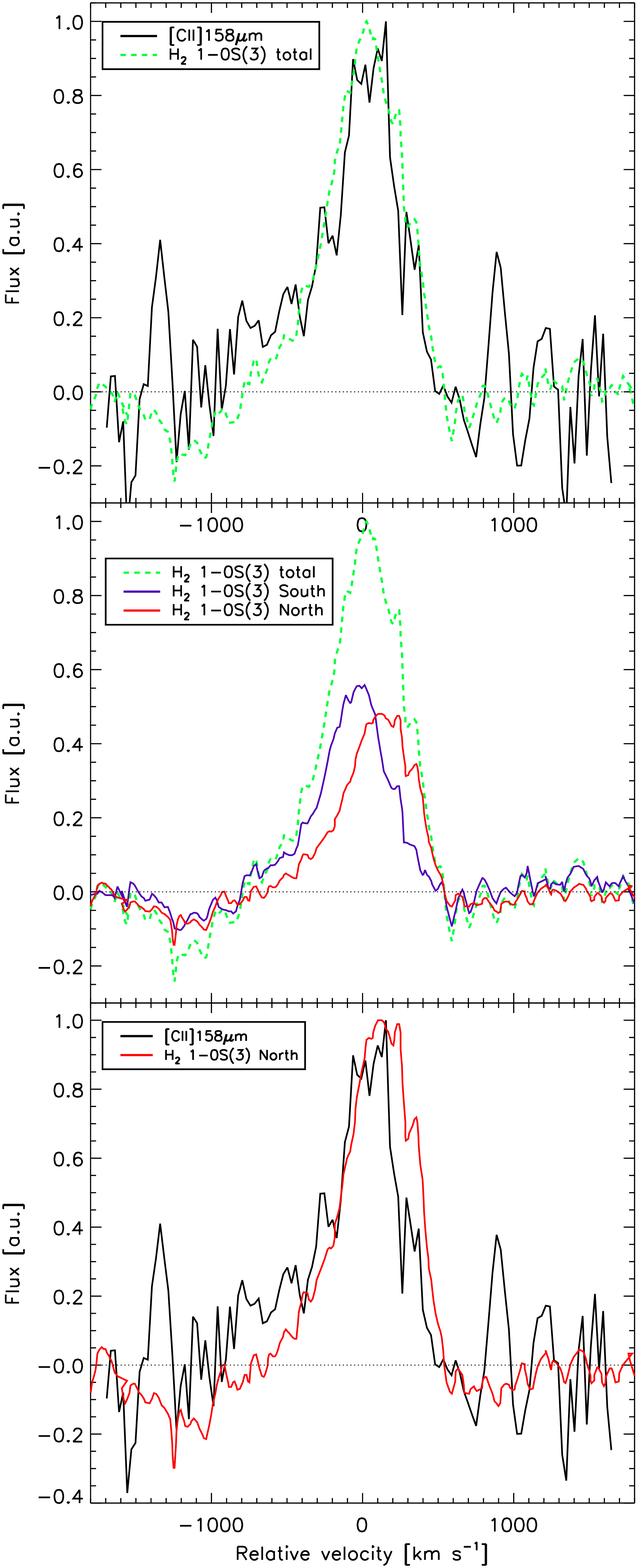}
\caption{ 
Comparison of the [C{\sc ii}]$\lambda$158$\,\mu$m and H$_2$ (1-0)~S(3)
line profiles in 3C~326N. All lines have been arbitrarily scaled (maximum line flux of 1) to
facilitate a direct comparison. 
\textit{Top panel:} Comparison between [C{\sc ii}] and the 
integrated H$_2$ 1-0S(3) 1.9576$\,\mu$m line profile (green dashed line). 
\textit{Central panel:} Total integrated H$_2$ 1-0S(3) line and decomposition of the 
H$_2$  line emission originating from the Northern disk hemisphere (red solid line) and
 the southern hemisphere (solid blue line), respectively. 
\textit{Bottom panel:}
The integrated [C{\sc ii}] line profile (solid black line), and H$_2$
line extracted from the Northern hemisphere (solid red line).
 }
\label{fig:h2c2}
\end{figure}

\begin{figure}
\centering
\includegraphics[width=0.5\textwidth]{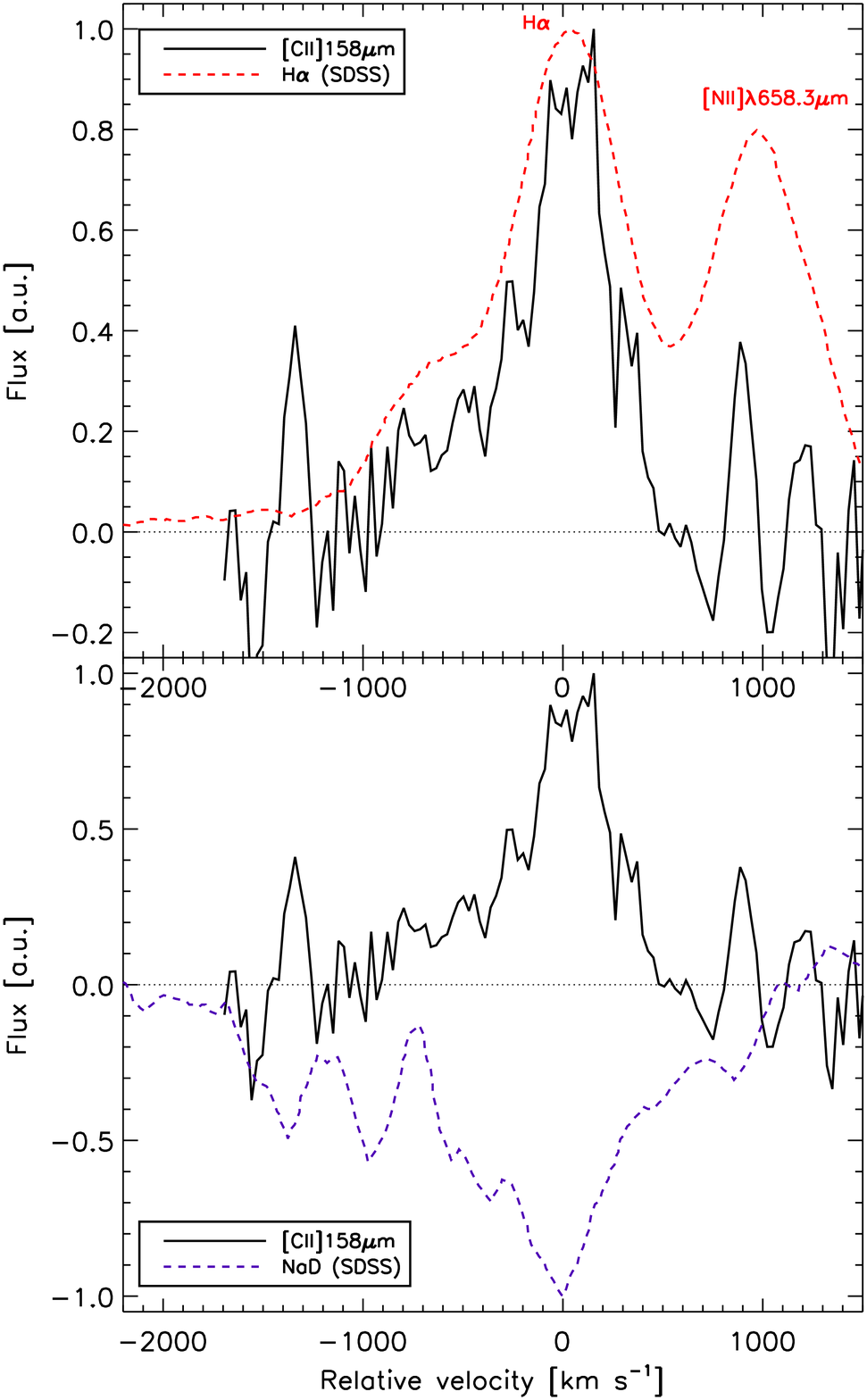}
\caption{
 {\it Top panel:} Comparison of the [C{\sc ii}]$\lambda$158$\,\mu$m
(solid black line) and H$\alpha$ (dashed red line) line profiles in 3C~326N. The
second line to the right of the H$\alpha$ line is [N{\sc ii}]$\lambda$6583\AA. 
{\it Bottom panel:} [C{\sc ii}] (solid black line) and Na~D absorption-line (dashed blue)
profiles. The maximum flux of each spectrum in emission has been normalized to 1 to ease the comparison, and 
the minimum absorption NaD flux has been normalized to -1.
}
\label{fig:hiic2}
\end{figure}

In Fig.~\ref{fig:c2co}, we compare the [C{\sc ii}] line to
the CO(1-0) profile obtained with the IRAM PdBI \citep{Nesvadba2010}.
[C{\sc ii}] and CO are the two tracers of the bulk of the cold gas mass for which we
have kinematic information in 3C~326N.  The
[C{\sc ii}] spectrum has a much better signal-to-noise ratio than the
CO(1-0) detection, which allows us to  constrain the cold gas
kinematics in 3C~326N better. 
The [C{\sc ii}] line seems broader than the CO(1-0) line, but unfortunately 
the quality of the CO(1-0) data is too low 
to do a detailed comparison between these two line profiles.  

The only spatially resolved kinematic information we have for
the molecular gas in 3C~326N is from the observation of the ro-vibrational
lines of H$_2$ with SINFONI on the VLT \citep{Nesvadba2011a}. Unlike
[C{\sc ii}] and CO(1-0), those H$_2$ lines trace a tiny fraction of the
molecular gas mass (at most a few percent). Figure~\ref{fig:h2c2} shows
a comparison of the H$_2$ 1$-$0 S(3) 1.9576$\,\mu$m line and the [C{\sc ii}] line
profiles. The top panel of Fig.~\ref{fig:h2c2}
shows that the integrated [C{\sc ii}] and 1-0~S(3)~H$_2$
line profiles look remarkably similar, given that the fractions of the gas mass probed
by these two lines are very different. However,  there are some small differences
worth noting.

The [C{\sc ii}]
blue wing is more pronounced than the H$_2$ line (see top panel
of Figure~\ref{fig:h2c2}). The analysis of the H$_2$ line kinematics shows that 
it has two main velocity components associated
with the southern (blueshifted peak) and northern (redshifted peak) disk
hemispheres. This decomposition is shown in the central panel of Figure~\ref{fig:h2c2}). 
The analysis of the H$_2$ velocity map reveals a
rotation pattern with a deprojected circular velocity of $290$~km~s$^{-1}$
\citep{Nesvadba2011a}.  Single Gaussian fits to these H$_2$ line profiles show prominent blueshifted wings in both hemispheres, with the strongest wing exhibited in the northern hemisphere. To facilitate the comparison, the bottom panel of Figure~\ref{fig:h2c2} 
shows the integrated [C{\sc ii}] and the H$_2$ line extracted over the northern disk hemisphere.

This comparison of the line kinematics suggests that the northern disk hemisphere makes  
a stronger contribution to the [C{\sc ii}] blue wing than the southern hemisphere. 
The reason for this is not clear. One explanation could be
that the turbulence is higher in the northern hemisphere of the disk. Indeed, we 
actually observe the brightest 1-0S(3) H$_2$ line flux
and the broadest line widths \citep[FWHM of $500-650$~km~s$^{-1}$ compared
to $350-450$~km~s$^{-1}$ in the southern hemisphere][]{Nesvadba2011a}. So
the [C{\sc ii}] line, like H$_2$, could be 
enhanced mainly by turbulent dissipation in the northern hemisphere of the
disk. This could also explain the larger [C{\sc ii}]/CO(1-0) ratio 
in the [C{\sc ii}] line at $\approx +250$~km~s$^{-1}$ (see
Fig.~\ref{fig:c2co}). This has to be confirmed with higher
angular resolution and higher signal-to-noise CO data.

The presence of a wind of neutral gas  in 3C~326N has been suggested
by an analysis of the Na~D absorption line profile \citep{Nesvadba2010}.
This outflowing gas likely contributes to the non-gaussianity of
the line profiles.  Fig.~\ref{fig:hiic2} shows that the [C{\sc ii}]
line profile match the Na~D absorption profile and the H$\alpha$
line emission, which both also show significant blueshifted wings. 
The ranges of negative velocities covered by the blueshifted wings in H$_2$ and in [C{\sc ii}] are very similar. However, the Na~D blue wing extends to higher velocities (up to $-1700$~km~s$^{-1}$) than the blue wings of the [C{\sc ii}] or H$_2$ profiles.

\section{[C{\sc ii}] as tracer of turbulent molecular gas}
\label{sec:diagnostics}

The [C{\sc ii}] line flux in 3C~326N is twice that of the 0-0~S(1)
17$\,\mu$m H$_2$ line. The [C{\sc ii}] line is remarkably bright, in comparison 
with the faint IR continuum flux of 3C~326N \citep[see][and \S~\ref{subsec:dustmodel}]{Ogle2007}. 
We find a [C{\sc ii}]-to-far infrared ratio of $0.035 \pm 0.014$ (see
below) and a [C{\sc ii}]-to-PAH$\,7.7\,\mu$m flux ratio greater than
2.3 \citep{Guillard2013}.
It provides, as we argue, evidence that most of the [C{\sc ii}]
line emission originates in the diffuse H$_2$ gas, with densities $n_{\rm H}$ up 
to a few $10^3$~cm$^{-3}$, in which a substantial
amount of turbulent energy is being dissipated.

\subsection{Mass of the [C{\sc ii}] emitting gas}
\label{subsec:CIIcoolrate}

The total mass of [C{\sc ii}] emitting gas (including Helium), $M_{\rm tot}^{\rm C^+}(\rm H)$, can be
estimated from the observed [C{\sc ii}] line luminosity\footnote{We assume that the [C{\sc ii}] emission is optically thin, which is justified later in Section~\ref{sec:outflow_turbulence}.}, $L_{C^+}$
[erg~s$^{-1}$], and the C$^+$ cooling rate per H atom, $\lambda _{C^+}$
[erg~s$^{-1}$~H$^{-1}$]:
\begin{equation}
\label{eq_MCII}
M_{\rm tot}^{\rm C^+}(\rm H) = \mu \, m_{\rm H} \frac{L_{\rm C^+}}{\lambda _{\rm C^+}} \, ,
\end{equation}
where $m_{\rm H}$ is the Hydrogen atom mass, and $\mu = 2.33$ the mean molecular weight. 
The cooling rate
from the single C$^+$ fine structure transition can be written as
\citep{Goldsmith1978}:

\begin{equation}
\label{eq_lambdaCII}
\lambda _{\rm C^+} = x_{\rm C^+} A_{ul} E_{ul} \left[ 1 + \dfrac{1}{2} e^{T_{ul}/T_{kin}} \left( 1 + \dfrac{A_{ul}}{C_{ul}} \right)  \right]^{-1} \, ,
\end{equation}
where $x_{\rm C^+} = [C^+]/[H]$ is the $C^+$ abundance,
$A_{u}=2.36\times 10^{-6}\,$s$^{-1}$ is the spontaneous emission rate,
and $E_{u} = 1.26\times 10^{-14}\,$erg is the excitation energy of the
upper $J=3/2$ level above the ground state, corresponding to an equivalent
temperature $T_{u} = E_{u}  / k = 91.25\,$K. The collision rate $C_{u}$
[s$^{-1}$] is the sum of the collision rates of the three main partners,
$e^-$, H and H$_2$:
\begin{equation}
C_{u} = R_{u}({\rm H_2}) n({\rm H_2}) +  R_{u}({\rm H}) n({\rm H}) + R_{u}(e^-) n(e^-) \, ,
\end{equation}
where the expressions of the downward rate coefficients $R_{u}$
[cm$^3$~s$^{-1}$] are reviewed in \citet{Goldsmith2012}.

We compute $\lambda _{C^+}$ at the critical density of the H$_2$ collision
partner, n$_{\rm H} = n_{\rm crit, H_2} = 6000\,$cm$^{-3}$, for a
molecular gas fraction of 90\% ($f_{\rm H_2} = 0.9$), so $n({\rm H_2}) = 0.9 \times n_{\rm H}
/ 2$, and $n({\rm H})  = 0.1\times n_{\rm H}$. This is justified because the total gas mass 
derived from the modeling of the dust far-infrared emission is similar to 
the molecular gas mass derived from rotational H$_2$ and CO(1-0) line 
measurements, so the molecular gas makes most of the gas mass 
(see Sect.~\ref{subsec:dustmodel} for a justification). There is no H{\sc i}
spectroscopy available for 3C~326N, so we caution that the molecular gas fraction is
not directly constrained from observations well, but this does not affect our
conclusions since the [C{\sc ii}] cooling rate does not depend much 
on the H$_2$ to H{\sc i} ratio (see below).  We assume a solar abundance for C and
assume that 40\% of the Carbon is in the gas phase, so $x_{\rm C^+} = 1.3\times
10^{-4}$, as in the Milky Way \citep{Cardelli1996}. The 3C~326N radio galaxy 
is likely to have a somewhat higher Carbon abundance, owing to its
higher mass\footnote{From the mass-metallicity relationship, a $3\times
10^{11}$~M$_{\star}$ galaxy like 3C~326N should have a metallicity
that is supersolar by about 0.4 dex.} and, to a lesser extent, to the
[$\alpha$/Fe] enhancement in stellar photospheres, but this is difficult
to quantify with the data we have at hand, and it does not affect our result
in any significant way. We estimate the electron density from the $C^+$
abundance, so $n(e^-) = x_{\rm C} n_{\rm H}$.

For instance, at a gas kinetic temperature of $T_{kin} = 100\,$K and n$_{\rm H} = n_{\rm crit, H_2} = 6000\,$cm$^{-3}$, we find  $\lambda _{\rm C^+} = 1.0 \times
10^{-24}$ and $1.4 \times 10^{-24}$~erg~s$^{-1}$~H$^{-1}$, respectively, for molecular fractions $f_{\rm H_2} = 0.9$ and 0.3. This confirms that the [C{\sc ii}] excitation depends weakly on the fraction of the gas that is molecular. 
The observed [C{\sc ii}] line luminosity
in 3C~326N is $L_{C^+} = 2.7 \times 10^{41}$~erg~s$^{-1}$ (Table~\ref{tab:spectrum}), which corresponds to a mass $M_{\rm tot}^{\rm C^+}(\rm H) = 4.5 \times 10^8$~M$_{\odot}$ (Eq.~\ref{eq_MCII}). 
Since
we calculated this mass at the H$_2$ critical density, the mass of
[C{\sc ii}] emitting gas in 3C~326N is likely to be higher. For the average physical 
conditions of the [C{\sc ii}]-emitting gas in 3C~326N, $T_{kin} \approx 100\,$K and 
n$_{\rm H} \approx 1000\,$cm$^{-3}$ (see Sect.~\ref{subsec:physicalconditions} and Figure~\ref{fig:CII_OI}
for a justification), we find $M_{\rm tot}^{\rm C^+}(\rm H) = 9.5\times
10^8$~M$_{\odot}$.

Could the [C{\sc ii}] line emission be attributed to ionized gas in 3C~326N? The mass of ionized gas in 3C~326N estimated from the H$\alpha$
emission-line luminosity is $M_{\rm HII} = 2\times 10^7\,$M$_\odot$
\citep{Nesvadba2010}. At $T_{kin} = 10^4$~K and n$_{\rm H} =
50\,$cm$^{-3}$ (i.e., the critical density at this temperature by collision
with $e^-$), the  [C{\sc ii}]  cooling rate is $\lambda _{\rm C^+}^{(WIM)}
= 3.1\times 10^{-25}$~erg~s$^{-1}$~H$^{-1}$, so the contribution of the
warm ionized medium (WIM) to the [C{\sc ii}] line cooling is
\begin{equation}
 L_{C^+}^{(WIM)} = \frac{M_{\rm HII}}{m_{\rm H}} \lambda _{\rm C^+}^{(WIM)} =  7.4 \times 10^{39} \ \rm erg \, s^{-1} \, .
\end{equation}
This is a factor 30 below the observed [C{\sc ii}] line luminosity
in 3C~326~N, so the [C{\sc ii}] emission cannot be accounted for by
recombination in the warm ionized medium.  

In 3C~326N, the total molecular gas
mass derived from the \mbox{CO(1-0)} line measurement is $M_{\rm tot}(\rm H_2) =
2 \times 10^9\,$M$_\odot$ (including He) for a Milky-Way-like
CO-to-H$_2$ conversion factor \citep{Nesvadba2010}. 
We conclude that the [C{\sc ii}] line emission traces about 30\% to 50\% of the total molecular gas mass derived from the CO(1-0) line, depending on the molecular gas fraction of the  [C{\sc ii}]-emitting gas. Since the masses of the other gas phases are lower than that estimate, 
 the molecular gas is the only gas reservoir that
can account for the observed [C{\sc ii}] line emission. 
A similar conclusion has been reached for the turbulent molecular gas present in the shocked intergalactic environment of Stephan's Quintet \citep{Appleton2013}.

\subsection{Dust and gas masses}
\label{subsec:dustmodel}

\begin{figure}
\centering
\includegraphics[width=0.5\textwidth,angle=0]{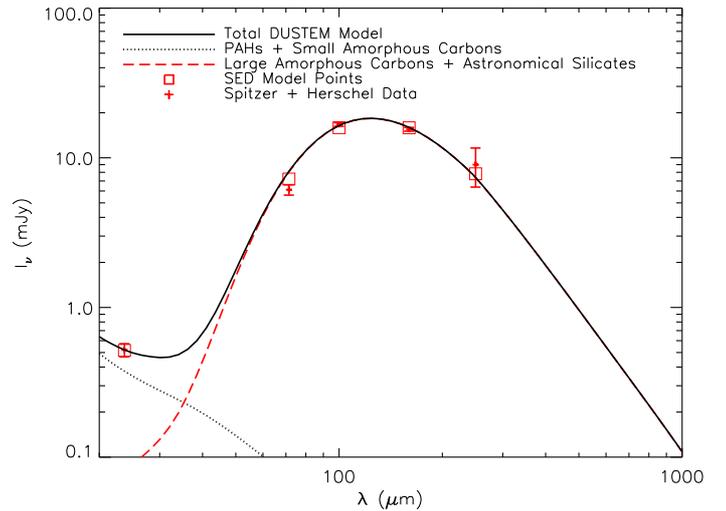}\\
\caption{3C~326N infrared spectral energy distribution (SED) fitted with
the DustEm model. The blue points show the photometric data, composed of MIPS, PACS and SPIRE flux measurements (Table~\ref{tab:FIR_fluxes}). The
solid black line is the best fit DustEm model, and the dash and dotted lines show the model spectra for the two main dust grain populations used in the fit. The
red squares show the photometric points computed from the model
SED. Color corrections given by the instruments manuals have been applied to homogenize the
different flux conventions between the instruments.  The fitting routine and model SED points take the bandwidth and transmission filters of the different instruments into account (see text for details). The DustEm fit parameters used in this paper are $G_0 = 9 \pm 1$ and a dust mass of M$_{\rm dust} = 1.1 \times 10^7\,$M$_\odot$.}
\label{fig:dustem_fit}
\end{figure}

\begin{figure}
\centering
\includegraphics[width=0.5\textwidth,angle=0]{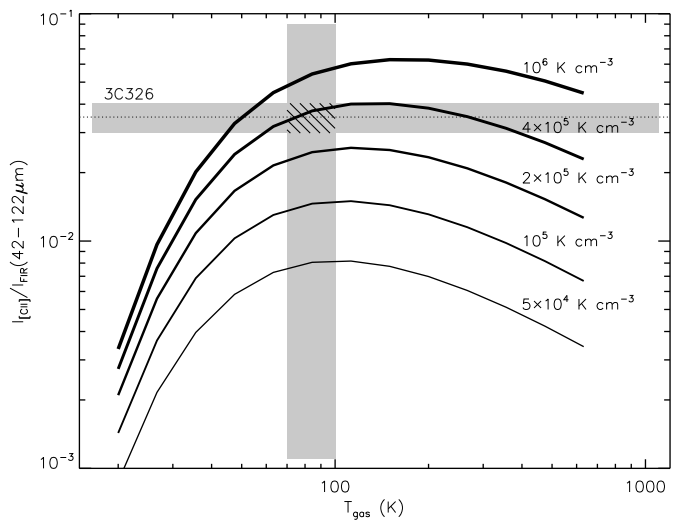}
\caption{
[C{\sc ii}]$\lambda \, 158\mu$m to far-infrared (FIR) intensity ratio as
a function of the gas temperature for different constant gas thermal pressures. The
[C{\sc ii}] cooling is computed from Eq.~\ref{eq_lambdaCII} and the FIR flux from 
the integration of our dust model over the range $42 - 122\,\mu$m (with a radiation field $G_0 = 9$, see
\S~\ref{subsec:dustmodel} and Fig.~\ref{fig:dustem_fit}) . The
observed ratio for 3C~326N, $I_{[CII]}/I_{FIR}(42-122\mu \rm m) = 0.035 \pm 0.014$
is indicated by the horizontal dotted line and the grey area represents the uncertainty, dominated by 
the large error on the SPIRE measurements. 
The vertical grey bar displays the range of gas temperatures as constrained by observations of the [C{\sc ii}] to [O{\sc i}]63$\mu$m line ratio (see Fig.~\ref{fig:CII_OI} and Section~\ref{subsec:physicalconditions}).
For this range of gas temperatures of $70<T<100$~K, the
range of pressures that matches both observational constraints is $P/k \approx 7\times
10^4 - 2\times 10^5$~K~cm$^{-3}$.
}
\label{fig:CIIdustvspressure}
\end{figure}

\begin{figure}
\centering
\includegraphics[width=0.5\textwidth,angle=0]{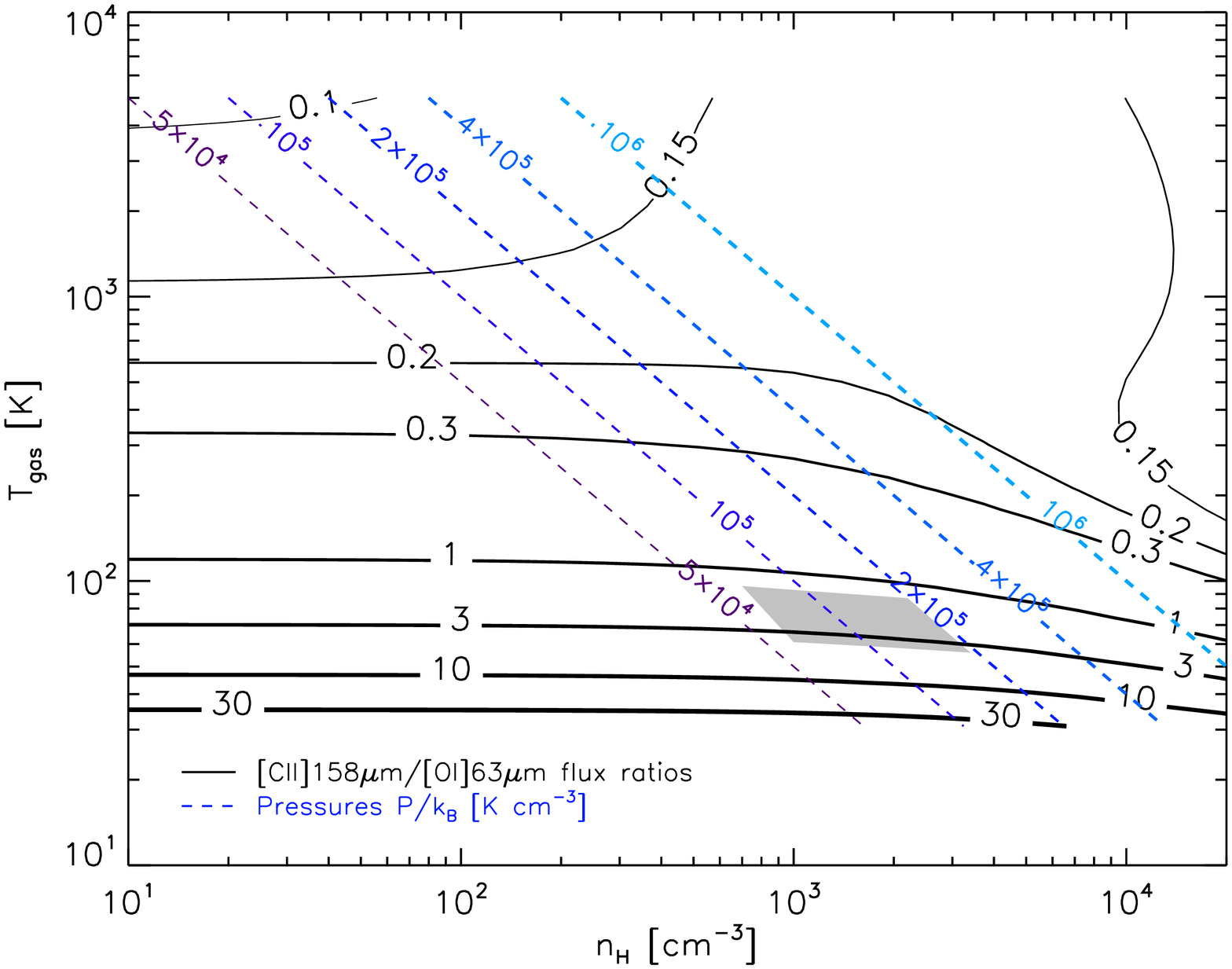}
\caption{Collisional calculation of the ratio of the [C{\sc ii}]$\lambda \, 158\mu$m to [O{\sc i}]$\lambda \, 63\mu$m line intensity
ratio in the temperature versus density plane. The dashed lines indicate constant values of the gas thermal pressure. 
The observed [C{\sc ii}] to [O{\sc i}]63$\mu$m ratio in 3C~326N [C{\sc
ii}]/[O{\sc i}]$= 2.4_{-0.4}^{+1.5}$ (see text for details). For the pressures inferred in 
Fig.~\ref{fig:CIIdustvspressure} and the constraint on the observed [C{\sc ii}] to [O{\sc i}]63$\mu$m ratio, the ranges of gas 
densities and temperatures compatible with the observations are  
$7 \times 10^2 < n_{\rm H} < 3\times 10^3$~cm$^{-3}$ and $70 < T < 100$~K respectively. The grey parallelogram gives the observational 
 constraints from Fig.~\ref{fig:CIIdustvspressure} and the [C{\sc ii}]/[O{\sc i}]ratio.
}
\label{fig:CII_OI}
\end{figure}

To estimate the total dust mass, we fit the infrared spectral energy distribution (SED) 
(MIPS+PACS+SPIRE) of 3C~326N with the DustEm dust model 
\citep{Compiegne2011}, using the IDL DustEm wrapper\footnote{Available
at \url{http://dustemwrap.irap.omp.eu/}}.  The measured FIR fluxes are
given in Table~\ref{tab:FIR_fluxes}. The only free parameters are the
radiation field intensity, $G_0$, and the different natures of dust grains. We use Galactic mass fractions of
the dust grains populations. This may not be entirely correct for such
a powerful radio galaxy, but this assumption does not significantly
affect our estimate of the total mass since we are mainly focusing on
the FIR emission coming from large grains. The results are
shown on Figure~\ref{fig:dustem_fit}, where we focus on the FIR part
of the SED because we are only interested here in the total dust mass.

From the DustEm fit we constrain the radiation field to be $G_0 = 9 \pm 1$, and a total
dust mass of M$_{\rm dust} = 1.1 \times 10^7\,$M$_\odot$. For a
dust-to-gas mass ratio of 0.007, this translates into a total gas mass of
M$_{gas}=1.6\times 10^9\,$M$_\odot$. Despite the very low star-formation
rate in 3C~326N \citep[$\le$0.07 M$_{\odot}$ yr$^{-1}$ from PAH and
24$\,\mu$m measurements,][]{Ogle2007}, this radiation field is of the order of magnitude of the UV field derived from GALEX measurements \citep[$G_0 = 6_{-2}^{+4}$,][]{Ogle2010}, and is mainly due to the old stellar population. 
We compared our DustEm modeling with the \citet{Draine2007} dust model. If
we adopt $G_0 = 9$, we find $M_{\rm tot}(\rm H_2) = 1.2 \times 10^9\,$M$_\odot$. 
For comparison, had we adopted $G_0 = 2$, we would have found
$M_{gas} = 2.9\times 10^9$ M$_\odot$. This shows that the total gas mass 
inferred from the modeling of the dust emission does not
depend strongly on $G_0$.

This total mass of gas derived from the modeling of the dust far-infrared spectral 
energy distribution is roughly in agreement with the mass of  warm  H$_2$ gas inferred from the fit of the pure-rotational H$_2$ excitation diagram \citep[$2.7\times
10^9$ M$_\odot$, including He,][]{Ogle2010}, and matches 
the mass of molecular gas derived from the CO(1-0) line measurement
\citep[2$\times 10^9\,$M$_\odot$][]{Nesvadba2010}. This good
correspondence  indicates that most of the gas mass in 3C~326N is molecular.

\subsection{Constraints on the physical state of the gas: [C{\sc ii}] as a probe of the warm, low pressure molecular gas}
\label{subsec:physicalconditions}

In this section we combine the [C{\sc ii}] cooling rate calculation
(\S~\ref{subsec:CIIcoolrate}) with the dust modeling
(\S~\ref{subsec:dustmodel}) to constrain the physical conditions
(pressure) of the gas required to account for the observed
[C{\sc ii}] luminosity.  We compute the ratio of [C{\sc ii}]
line emission to far-infrared dust emission as a function of
the thermal gas pressure, under the same assumptions given in
\S~\ref{subsec:CIIcoolrate}. We showed in Section~\ref{subsec:CIIcoolrate} that the [C{\sc ii}] line
emission traces mostly warm molecular gas. So again 
we compute the [C{\sc ii}] cooling rate assuming that 90\% of the gas is in molecular form, and 10\% is atomic, with an electron fraction dominated by Carbon ionization.
Fig.~\ref{fig:CIIdustvspressure} shows the expected ratio of the [C{\sc
ii}] to  far-infrared flux for a range of gas thermal pressures between $P/k=2.5 \times 10^4$~cm$^{-3}$~K and $4
\times 10^5$~cm$^{-3}$~K, and temperatures $T=10-1000$~K, with $P/kT = n_{\rm H}$. 
The far-infrared flux is computed from the dust model discussed 
in Sect.~\ref{subsec:dustmodel} and shown in Fig.~\ref{fig:dustem_fit}, 
after integration over the wavelength range 42-122$\,\mu$m \citep[as 
commonly used for extragalactic studies, e.g.,][]{Helou1988, Stacey2010}. 
We find a far-infrared luminosity of 
$L_{FIR}(42-122\,\mu m) = 2.0 \pm 0.8 \times 10^9$~L$_{\odot}$.
 The curves in Fig.~\ref{fig:CIIdustvspressure} peak at $\approx 100\,$K, 
 close to the excitation temperature of the
upper level $J=3/2$ of the transition. At this temperature, optimal for
[C{\sc ii}] excitation, the observed line-to-continuum ratio is matched
for a range of pressures $ 7 \times 10^4 < P/k < 2\times 10^5$ K~cm$^{-3}$.

For the total gas mass inferred from the modeling of the dust far-infrared 
continuum ($1.6\times 10^9\,$M$_\odot$), this corresponds to a [C{\sc ii}] cooling rate 
$\lambda_{\rm C^+} = 1.4\times 10^{-25}$ erg s$^{-1}\,$H$^{-1}$ for our  
estimate of $G_0 \approx 9$. This is a factor 5 higher than for the diffuse ISM in the Milky
Way \citep[$2.6\times 10^{-26}\,$erg s$^{-1}$~H$^{-1}$,][]{Bennett1994}.

To help in breaking the degeneracy between temperature and density, we
use the \textit{Herschel} PACS observations of the [O{\sc i}]$\lambda
63\mu$m line with  $F([{\rm OI}]) = 5.5_{-0.22}^{+0.07} \times 10^{-18} \,
$W~m$^{-2}$. The [O{\sc i}] data are part of a separate program -- P.I. Ogle --,
 and the spectrum will be published in a separate paper
as part of a larger sample of radio-galaxies \citep{Guillard2014a}. 
The uncertainty on the [C{\sc ii}]-to-[O{\sc i}] line flux ratio is dominated by the low signal-to-noise [O{\sc i}] detection.
Figure~\ref{fig:CII_OI} shows the results of collisional excitation
calculation of the [C{\sc ii}]-to-[O{\sc i}] line intensity ratio in the temperature versus density plane. The
cooling from [C{\sc ii}] is computed from Eq.~\ref{eq_lambdaCII} and for
the [O{\sc i}] cooling we solve the three-level fine-structure O($^3P$)
oxygen system. We considered collisions with H, H$_2$ and electrons.
We use the O($^3P$)-H collisional rate coefficients computed by
\citet{Abrahamsson2007} which we fitted as a function of the temperature
to get analytical expressions.  The rate coefficients for collisions
with H$_2$ are taken from \citet{Jaquet1992}, and those with electrons
are from \citet{Bell1998}.  
These calculations show that the gas temperature is well constrained by the [C{\sc ii}]-to-[O{\sc i}] line ratio. 
The comparison of these calculations with
the observed [C{\sc ii}]-to-[O{\sc i}] ratio shows that, for the range of pressures 
matching the observed [C{\sc ii}]-to-FIR ratio (Fig.~\ref{fig:CIIdustvspressure}), 
the average density of the gas has to be relatively low, within the range 
 $7\times 10^2 < n_{\rm H} < 3\times 10^3$~cm$^{-3}$ (Fig.~\ref{fig:CII_OI}).

 In addition, the fitting of the rotational H$_2$ line emission with a thermal 
equilibrium model shows that the bulk of the warm H$_2$ gas 
must be at a temperature of $\approx 100$~K to match the observed luminosity 
in the low-J H$_2$ rotational lines, i.e., 0-0$\,$S(0) and 0-0$\,$S(1), 
\citep[e.g.,][]{Ogle2010, Guillard2012}.  Both constraints
on the warm H$_2$ density ($7\times 10^2 < n_{\rm H} < 3\times 10^3$~cm$^{-3}$) and
temperature ($70 < T< 100$~K) rule out the possibility that the [C{\sc
ii}] emission is coming from a lower fraction of the gas mass at higher
pressure, because the [C{\sc ii}]-to-FIR flux ratio is dropping as
the temperature deviates from 100~K (Fig.~\ref{fig:CIIdustvspressure}).
Furthermore, since dense, high-pressure gas would be luminous in the
CO lines, further constraints on the gas density can be obtained from
CO observations.  A preliminary inspection of the SPIRE FTS spectrum
of 3C~326N shows no detection of high-J CO lines (P. Ogle, priv. com.),
which seems to confirm that the molecular gas has a low average density or a low filling fraction of dense gas.
We conclude that the average pressure of the warm H$_2$ gas in 3C~326N is
$P/k \approx 10^5\,$K~cm$^{-3}$, with a temperature of $T \approx 100\,$K
and a density $n_{\rm H} \approx 10^3$~cm$^{-3}$.  Therefore, in 3C~326N, 
the mid-IR rotational lines of H$_2$ and  the [C{\sc ii}] line
trace the same warm molecular gas. 

\section{What is the heating source of the gas?}
\label{sec:gasheating}

\subsection{An extreme [C{\sc ii}] line luminosity}

In Sect.~\ref{subsec:physicalconditions}, we computed a [C{\sc ii}] to FIR
flux ratio of $I_{[CII]}/I_{FIR}(42-122\mu \rm m)
= 0.035 \pm 0.014$, see Fig.~\ref{fig:CIIdustvspressure}. This is unusually high, a factor of 3 to 5 more than what is observed in typical star-forming galaxies \citep{Malhotra2001, Stacey2010, Diaz-Santos2013, Rigopoulou2014, Herrera-Camus2014}. This ratio is very similar to what is found in the shocked intergalactic medium of the Stephan's Quintet galaxy collision \citep[0.04--0.08,][]{Appleton2013}, and, to a lower extent, in the HCG~57 group \citep[up to 0.017,][]{Alatalo2014a}.  As we see in this section, the [C{\sc ii}] emission in 3C~326N cannot be powered by star formation.

We demonstrated in Section~\ref{sec:diagnostics} that the [C{\sc ii}] line luminosity in 3C~326N implies that a 
significant fraction (30--50\%) of the molecular gas  mass 
($5.5-9.5 \times 10^8$~M$_{\odot}$) has to be
warm ($70 < T < 100$~K). In the
following sections, we discuss the potential heating sources of this gas.

\citet{Ogle2010} and \citet{Nesvadba2010} ruled out UV and X-rays as possible
sources for the heating of the warm H$_2$ gas in 3C~326N. Since
we demonstrated that most of the [C{\sc ii}] line emission is
coming from the warm molecular gas, this is a fortiori
valid for the [C{\sc ii}] excitation. 
Furthermore, the [C{\sc ii}] to PAH luminosity ratio is too high 
([C{\sc ii}] / PAH$\,7.7\,\mu$m$ > 2.5$) to be accounted for solely by 
photoelectric heating \citep{Guillard2013}, which
rules out photon heating as the main source for the [C{\sc ii}] emission.
Therefore, the two remaining possible heating sources of the gas in
3C~326N are the turbulent heating and the cosmic rays.

\subsection{Turbulent heating}
\label{subsec:turb_dissipation}

Dissipation of turbulent energy (mechanical heating) has a strong impact on the excitation conditions and physical state of the diffuse ISM and molecular gas in galaxies \citep[e.g.,][]{Godard2014, Kazandjian2012, Rosenberg2014}.
\citet{Nesvadba2010} argue that the dissipation of mechanical energy
is the most likely excitation source of the molecular gas in 3C~326N. 
This turbulence must be sustained by a vigorous energy source because 
the radiative cooling time of the warm molecular gas is short 
\citep[$\approx 10^4$~yr;][]{Guillard2009}.  
They demonstrate that a few percent of the jet kinetic luminosity
(at least a few $10^{44}$~erg~s$^{-1}$) deposited into the gas is
sufficient to drive the outflow and power the observed H$_2$ and H{\sc ii}
line luminosity, which makes jet kinetic energy a plausible source for
the heating of the molecular gas.

Within this framework, [C{\sc ii}] line observations allow us
to build a more complete mass and energy budget. In particular, we
update the estimate of the turbulent gas velocity dispersion $\sigma
_{turb}$ required to balance the total [C{\sc ii}] + H$_2$ line cooling
rate. Turbulent heating is energetically possible if the total turbulent kinetic luminosity is greater than the observed line luminosity:
\begin{equation}
\label{eq:turb_heat_rate}
 \dfrac{3}{2} M_{\rm tot}({\rm H_2})  \frac{\sigma _{turb} ^3}{H_g} > L_{\rm [CII]} + L_{\rm H_2} \ ,
\end{equation}
where $M_{\rm tot}(\rm H_2) = 2 \times 10^9$~M$_{\odot}$ is the total
molecular gas mass, $\sigma _{turb}$ is the turbulent velocity dispersion,  and $H_g$  the characteristic injection scale
of the turbulence, which we assume to be the vertical scale height
of the gas (see Sect.~\ref{subsec:scale_height} for a discussion of that assumption).  
The minimal velocity dispersion required to balance the line cooling is thus given by
\begin{equation}
\label{eq:min_vel_disp}
\sigma _{turb} > 90 \ \left( \frac{\rm L_{\rm [CII]+H_2} / M_{\rm tot}}{0.18 \ \rm L_{\odot}/M_{\odot}}\right)^{1/3} \left(  \frac{H_g}{1\, \mbox{kpc}} \right)^{1/3} \ \ \rm{km \ s^{-1}} \ .
\end{equation}
The observed [C{\sc ii}]$\lambda 158\,\mu$m
line cooling rate per unit mass in 3C~326N is $L_{\rm [CII]} / M_{\rm tot}(\rm H_2)
= 0.044 \pm 0.006~ \rm L_{\odot} / M_{\odot}$, and the H$_2$ rotational
line cooling rate is $L_{\rm H_2} / M_{\rm tot}(\rm H_2) = 0.14 \pm 0.02~ \rm
L_{\odot} / M_{\odot}$ \citep{Ogle2010}. 
In Eq.~\ref{eq:min_vel_disp} we chose to compute the minimal velocity dispersion for a relatively large vertical scale height, $H_g=1$~kpc. This scale height is  within the range of gas scale heights we constrain in sect.~\ref{subsec:scale_height} and is comparable to 
the radius of the molecular disk seen in ro-vibrational H$_2$ line emission \citep{Nesvadba2011a}. 
For those values of cooling rate and scale height, we obtain $\sigma _{turb} > 90$~km~s$^{-1}$. 

 Our [C{\sc ii}]
data, as well as near-IR H$_2$ measurements, show that the observed 
gas velocity dispersions are within the range 
$160 < \sigma < 350$~km~s$^{-1}$ (from 
the linewidth measurements of the individual velocity components). 
Based on the analysis of the
H$_2$ line velocity maps, \citet{Nesvadba2011a} estimate ratios of
circular-to-turbulent velocities $v_c / \sigma = 0.9 - 1.7$.  This ratio
is likely to be underestimated \citep[by $\approx$15\%,][]{Nesvadba2011a}
because of the effect of beam-smearing\footnote{The minor axis of
the H$_2$ disk seen in the near-IR lines is 0.55'' in radius, while
the spatial resolution is 0.7''.}. Beam smearing strongly affects the
determinations of the velocity structure in the data by transforming
what are in reality circular motions into apparent higher velocity
dispersions. Taking beam smearing into account, 
\citet{Nesvadba2011a} show that the contribution of the systematic rotation to 
the line width is $\sigma _{rot} \approx 100$~km~s$^{-1}$. By subtracting in quadrature the 
contribution of the rotation to the observed $\sigma$ from [C{\sc ii}] data, we find that the 
turbulent velocity dispersions in 3C~326N are within the range 
$120  < \sigma _{turb} < 330$~km~s$^{-1}$. 
Therefore, the energetic condition given above in Eq.~\ref{eq:turb_heat_rate} is satisfied, 
even for high values of the scale height. 
The turbulent kinetic luminosity
associated with the H$_2$ gas velocity dispersion alone on the physical
scale of the disk is at least a factor of 2 higher than the total [C{\sc ii}]+H$_2$
cooling rate.

As argued in previous papers \citep{Guillard2009, Nesvadba2010}, shocks are responsible for the dissipation of a large percentage of the total kinetic energy in H$_2$-luminous galaxies, but most of the kinetic energy dissipation 
does not occur within the ionized gas through dissociative shocks. It occurs mostly in non-dissociative, low-velocity C-shocks (velocities lower than 40~km~s$^{-1}$).
Since the kinetic energy is mostly dissipated in the molecular gas, molecular viscosity causes the turbulent dissipation. The timescale associated with the turbulent energy dissipation timescale is given by 
\begin{equation}
\label{eq:turb_diss_time}
\tau _{\rm diss} \approx 5\times 10^7 \left( \frac{\sigma _{turb}}{200\ \rm km \ s^{-1}} \right)^2 \left( \frac{0.18 \ \rm L_{\odot}/M_{\odot}}{\rm L_{\rm [CII]+H_2} / M_{\rm tot}(\rm H_2)} \right)\ \rm \ [yr].
\end{equation}
Given the range of turbulent velocity dispersion inferred above, the dissipation time ranges between $2\times 10^7$ and $1.5 \times 10^8$~yr. Therefore, the dissipation time is comparable to or longer than the lifetime of the radio jet or the duty cycle of the jet activity. The dissipation of turbulent energy in 3C~326N can power the line emission coming from the molecular gas between the times when the jet is active.

We also define $\epsilon$ as the ratio between the total [C{\sc ii}]+H$_2$ luminosity
and the turbulent heating rate. Within the framework given by the above condition 
(Eq.~\ref{eq:turb_heat_rate}), $\epsilon < 1$, 
which means that not all the turbulent energy dissipated is radiated into the  [C{\sc ii}] and H$_2$ lines.
Then Eq.~\ref{eq:turb_heat_rate} can be rewritten as a relationship between
the gas scale height and the gas velocity dispersion:
\begin{equation}
\label{eq:sca_hei_turb_diss}
H_g =  \frac{3}{2}\,  \epsilon \, \sigma _{turb} ^{3} \left( \frac{L_{\rm [CII]+H_2}}{M_{\rm tot}(\rm H_2)} \right)^{-1} \ .
\end{equation}
We  use this relation in sect.~\ref{subsec:scale_height} to discuss the 
dynamical state of the gas in the disk of 3C~326N.

\subsection{Heating by cosmic rays}
\label{subsec:cosmicrays}

We compute here the cosmic ray ionization rate required to balance the
[C{\sc ii}]$+$H$_2$ line cooling rate.  The gas cooling rate through
the [C{\sc ii}] line and the H$_2$ rotational lines of $4\pi\ (I(CII)
+ I(H_2)) / N_H= 4.4 \times 10^{-25}$ erg s$^{-1}$ H$^{-1}$. The
heating energy deposited per ionization is $\sim$13~eV for H$_2$
gas \citep{Glassgold2012}. Thus, to heat the gas, a cosmic ray
ionization rate of $\zeta_{\rm H_2} = 2 \times 10^{-14}$ s$^{-1}$ is needed.
This required ionization rate is a factor about 60 larger than in the
diffuse interstellar gas of the Milky Way \citep{Indriolo2012}, and a
factor of 18 greater than the value derived for the gas near the galactic
center \citep{Goto2013}.

However, each ionization produces two ions and one H$_2$ dissociation, so the gas has to
be dense enough to reform H$_2$ and be mostly molecular.  The formation rate ($\Gamma_{\rm
H_2} n({\rm HI}) n({\rm H})$) should balance the destruction rate $\zeta
/ n({\rm H_2})$ of H$_2$.  This balance is written $\Gamma_{\rm H_2} =
\zeta_{\rm H_2} f_{\rm H_2}/(1-f_{{\rm H_2}})/2/n({\rm H})$ where $f_{\rm H_2} =
2\ n({\rm H_2})/n({\rm H})$ is the molecular fraction of the gas.  As a result,
the condition for the gas to be predominately molecular can be written as
\begin{equation}
n_{\rm H} \geqslant \frac{\zeta_{\rm H_2} \times f_{\rm H_2}}{\Gamma_{\rm H_2}
\times 2 (1 - f_{\rm H_2})} = 3.3 \times 10^2 \frac{f_{\rm H_2}}{1 -
f_{\rm H_2}}\ \rm [cm^{-3}] \, , 
\end{equation}
 if we adopt the canonical value of
$\Gamma_{\rm H_2} \approx 3\times 10^{-17}$~cm$^3$ s$^{-1}$ \citep[e.g.,][]{Gry2002, Habart2004},
and the cosmic ray ionization rate derived above ($\zeta = 2 \times
10^{-14}$ s$^{-1}$). For a gas that is mostly molecular ($f_{\rm H_2}
= 0.9$), we find that $n_{\rm H} \geqslant 3 \times 10^3\,$cm$^{-3}$. 
This lower limit on the gas density corresponds to the upper 
boundary of the gas density range derived from the observed  [C{\sc ii}]-to-[O{\sc i}] line ratio 
(Fig.~\ref{fig:CII_OI}). 

The average gas density derived from [C{\sc ii}], [O{\sc i}], 
and FIR continuum photometry ($7 \times 10^2 < n_{\rm H} < 3\times 10^3$~cm$^{-3}$, see
Sect.~\ref{subsec:physicalconditions}) does not permit a high H$_2$
re-formation rate and thus these conditions are not favorable for
keeping the gas molecular in the presence of intense cosmic ray radiation.
We therefore do not favor cosmic rays for playing a dominant role
in the heating and excitation of the molecular gas.  However,
the lower limit on the gas density derived above is close enough 
to the estimated range of densities (Fig.~\ref{fig:CII_OI}) that it is not possible to conclusively rule out
the cosmic ray heating, provided that such a high ionization rate can
be maintained on the large scale of the galaxy's interstellar medium.
We do not have strong constraints on
the diffuse cosmic ray intensity in radio galaxies, which may be enhanced
if the jets flare \citep{Biermann2012, Laing2013, Meli2013} but could also be
low in the absence of star formation as is observed in 3C~326N.

Because they penetrate deeper than UV photons into molecular clouds,
cosmic rays may play an important chemical role in dissociating CO
molecules (via secondary photons), thus maintaining a high $\rm C^+/CO$
fractional abundance. \citet{Mashian2013} have indeed shown that  in 
$n_{\rm H_2} =  10^3$~cm$^{-3}$, $T=160$~K, UV-shielded gas, the gas can be
molecular ($1 < n_{\rm H_2}/n_{\rm H}<50$) with $n(C^+)/n(CO) \approx 0.8$
for $ 3\times 10^{-16} < \zeta < 10^{-14}$~s$^{-1}$.  While cosmic rays
may not play a dominant role in heating and exciting the molecular gas,
they very likely play an important role in the chemistry of the gas.

\section{Outflow and turbulence in 3C~326N}
\label{sec:outflow_turbulence}

 In this section we discuss the two phenomena that can make the [C{\sc ii}] line profile very broad and skewed  (see Fig.~\ref{fig:spectrum}): a massive outflow and turbulence. These two processes are at work in 3C~326N and are coupled, since the outflow is likely to power some of the turbulence within all phases of the gas.

\subsection{The wind and outflow rates in 3C~326N}
\label{subsec:kinematics_wind}

Since the [C{\sc ii}] line emission traces $\approx 30$\% of the total molecular gas mass 
in 3C~326N (Sect.~\ref{sec:diagnostics}), associating the broad wing with an outflow component would
require that a large amount of the neutral medium is outflowing.  The
blueshifted wing accounts for 35\% of the total [C{\sc ii}] line flux
(Table~\ref{tab:spectrum}).  To estimate the fraction of the gas mass
that is actually outflowing and escaping from the galaxy, we integrate
the [C{\sc ii}] line profile from the highest observed velocity in the [C{\sc ii}] line 
($-1000$~km~s$^{-1}$) to the estimate of the escape velocity at the largest H$_2$ radius 
\citep[$-400$~km~s$^{-1}$,][]{Nesvadba2011a}.
 Assuming that the C$^+$ excitation and abundance are the same in the disk and in the outflow, 
 we find that this fraction is $11 \pm 2$~\%, corresponding to a total mass of outflowing gas of $1.8 \pm
0.3 \times 10^8$ M$_{\odot}$.

Using the upper limit of the size of the [C{\sc ii}] emitting region, the 10\arcsec\ PSF of PACS, 
and a Doppler parameter of 350 km s$^{-1}$ for the wind \citep{Nesvadba2010},
we estimate the dynamical time of the outflow to be lower than $10^7$ yrs.  This gives a lower 
limit for the outflow rate of $> 20$~M$_{\odot}$ yr$^{-1}$ and an upper limit on the depletion
time of $< 10^8$~yrs. Such a timescale is longer than or comparable to
the lifetime of the radio jet \citep[$10^7 - 10^8$~yrs,][]{Willis1978}
and comparable to the orbital timescale \citep[$\approx 3 \times
10^7$~yrs,][]{Nesvadba2011a}.

The outflow rate we estimate from the [C{\sc ii}] emission
is about half the mass outflow rate of neutral gas estimated from
Na~D absorption, 40$-$50 M$_{\odot}$ yr$^{-1}$ \citep{Nesvadba2010},
although the Na~D absorption extends to higher terminal velocities,
up to v$=-$1800 km s$^{-1}$, and is therefore likely to sample gas
that we do not detect in [C{\sc ii}].  The reason for this difference
may be either observational, in the sense that our PACS data do not have the
signal-to-noise ratio needed to probe the full velocity range of the
wind, or astrophysical, since the Na~D absorption line could probe gas down to lower
column densities, including gas that is more readily accelerated. 
If the mass outflow rate in 3C~326N is as high as $20-50$~M$_{\odot}$ yr$^{-1}$, 
this means that we are observing this galaxy at a very specific and 
brief time in its evolution, and one can wonder why we would 
observe a gas disk at all with such a short depletion timescale. 
Comparable mass outflow rates driven by AGN have been reported by, for example, 
\citet{Morganti2013}, \citet{McNamara2014} and \citet{Garcia-Burillo2014}. 
To replenish the disk, the inflow rate would need to be comparable to the outflow 
rate on at least the orbital timescale, which is comparable to the mass deposition rates 
observed in galaxy clusters \citep[e.g.,][]{Bregman2005a, Fabian2005} and which seems very high and unlikely for such an isolated galaxy. 
Therefore, we conclude that AGN with such high relative mass outflow rates are a transient phenomenon, 
which we are able to detect because radio galaxies (and quasars) are, by selection, 
caught in a very specific and short phase of their life. 
This result is in agreement with the low detection rate 
of the occurrence of H{\sc i} outflows seen in absorption \citep{Gereb2014}.

In 3C~326N, the [C{\sc ii}] line is unlikely to suffer
significantly from opacity effects\footnote{At velocity dispersions of
400~km~s$^{-1}$, the [C{\sc ii}] line would become optically thick at
column densities above $N_{\rm H} = 5 \times 10^{23}$~cm$^{-2}$, which is
more than two orders of  magnitude higher than the total column estimated
from H$_2$ spectroscopy \citep{Ogle2010}.}. If the broad wing component
of the [C{\sc ii}] line traces the cold outflowing gas, then why is the
line asymmetric?  The spatial distribution of the outflowing gas, and
the distribution of the filling factor of the cold gas are unlikely to
be homogeneous and isotropic, which could skew line profiles. Asymmetric
H{\sc i} absorption line profiles associated with off-nuclear outflows
are indeed observed in radio galaxies \citep[e.g.,][]{Morganti2005a, Gereb2014}.
In fact, the high degree of symmetry seen in the broad line profiles
towards e.g., high-redshift quasars \citep{Maiolino2012}, and often
interpreted as outflows, is remarkable.  The complex geometry of the
gas distribution also increases the uncertainty on the estimates of the
mass outflow rates.

\subsection{Line wings and turbulence on large scales}
\label{subsec:kinematics_turbulence}

Turbulence within the entire gas disk could also broaden and skew the [C{\sc ii}] line profile, making the line wing prominent. 
In addition to rotation and outflow, the line profiles in 3C~326N
may also trace the stochastic injection of turbulent energy on large scales 
by the radio jet into the gas. If this contribution is important, our estimates of mass outflow rates
could be significantly over-estimated \citep[see also][for the contribution of turbulence to the broad H{\sc i} absorption profile observed in 3C~293]{Mahony2013}.

 Because the largest eddies are rare in space and
time, they are not well statistically sampled by the observations, which
causes irregularities in the line profiles. In such a perspective where the
turbulence is driven by the radio jet, numerical simulations suggest that
the backflow of the jet can inject turbulent energy into the gas over
the vertical scale height of the disk \citep{Gaibler2012}. Therefore,
the largest turbulent cells are likely to be a few kpc in size, and the
number of correlation lengths is small, possibly explaining why
the spectral line profiles are jagged. This could mimic outflowing gas, but, 
because those velocity structures are short-lived and chaotic, the gas actually 
does not have the time to escape out of the galaxy.

The line profiles observed in 3C~326N are very reminiscent of what is 
observed in the spectral line shapes
(e.g., CO) of Galactic molecular clouds, where their irregularities
can be a signature of the statistical properties of turbulence
\citep{Falgarone1990, Falgarone1992}. 
Studies of the probability density distributions of the velocity field
(or vorticity, pressure, or energy dissipation rate) in the Galactic ISM show that their non-Gaussian behavior could in fact be a consequence
of stochasticity on small scales \citep[see][for observational evidence of intermittency]{Hily-Blant2008}.

\section{Dynamical state of the molecular gas in the 3C~326N disk}
\label{sec:dynamical_state}
\label{subsec:scale_height}

In this section we discuss the impact of turbulence on the dynamical state 
of the molecular gas and star formation. We 
investigate if our interpretation of the H$_2$ and [C{\sc ii}] line emission, 
as tracers of turbulent energy dissipation, fits with a plausible dynamical state of the 
molecular gas in 3C~326N. 

The hydrostatic equilibrium condition between the gas plus stellar gravity and the total (thermal plus turbulent) gas velocity dispersion determines the vertical
scale height of the disk \citep[e.g.,][]{Lockman1991, Malhotra1994}.
In 3C~326N, we expect that turbulent support of gas in the vertical
gravitational field is important.  In order to determine if the disk
is self-gravitating or not, we calculate the balance between the turbulent
pressure gradient and the gravity (in the vertical direction perpendicular to the disk plane) and we derive an expression for the vertical scale height.  Then we investigate if such an expression can be reconciled with the 
expression of the scale height constrained from the balance between 
the turbulent heating rate and the observed [C{\sc ii}]+H$_2$ cooling rate.

Because the jet is roughly perpendicular
to the orientation of the disk \citep{Rawlings1990}, it is likely that energy is being
transferred through turbulence generated by the over-pressured hot gas
created by the heating of the jet. This gas is also produced in models of relativistic jets
as they expand into the surrounding medium \citep[e.g.,][]{Gaibler2011,
Wagner2012}.  As in Sect.~\ref{eq:turb_heat_rate}, the scale of injection
of kinetic energy is comparable to the vertical scale height of the gas in the disk,
$H_g$.  This is a reasonable assumption as the Mpc-size of the radio emission
is much larger than the size of the disk.

The turbulent pressure can be written as $P_{turb} = \rho _g  \sigma _{g}
^2 = (\Sigma _{g}/ 2 H_g) \sigma _{g} ^2$, where $\rho _g$ is the average
mid-plane gas mass density, $ \Sigma _{\rm g}$  the gas mass surface
density, $H_g$  the gas vertical scale height, and $\sigma _{g}$ is the
vertical gas velocity dispersion\footnote{We use this velocity dispersion in the turbulent pressure formula because we write the vertical hydrostatic equilibrium condition}.  The pressure in the disk
plane due to the gas and stellar gravity can be written as $ P_{grav} =
(2 G)^{1/2} \Sigma_g \sigma _{g} \left( \rho_{\star}^{1/2} + (\frac{\pi}{4}
\rho _g)^{1/2} \right)$ \citep{Blitz2004, Ostriker2010}, where $\rho _{\star}$ is the
stellar mass density and $G$ the gravitational constant. The vertical 
equilibrium condition between the turbulent pressure gradient and the gravity
has two limiting forms.

\begin{figure}
\centering
\includegraphics[width=0.49\textwidth]{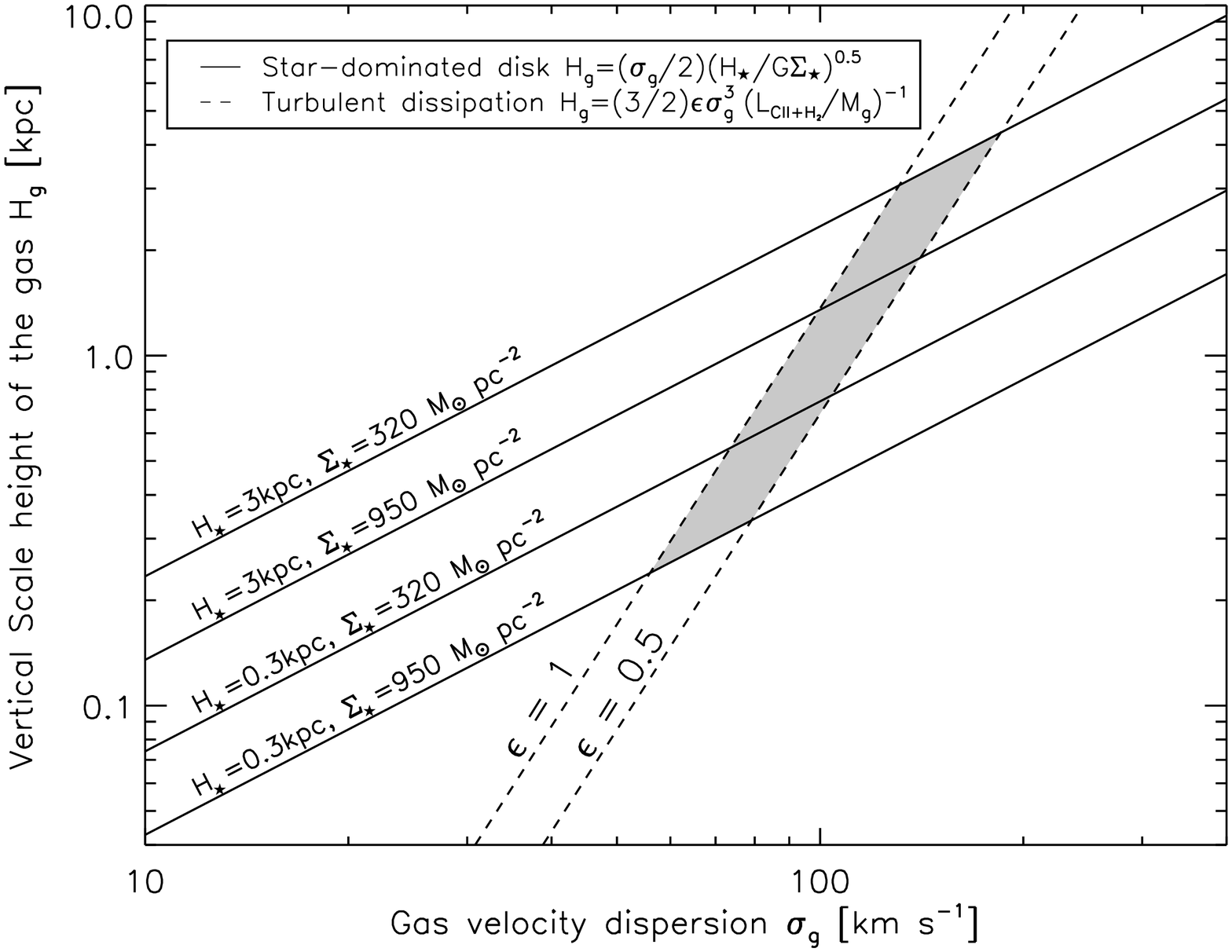}
\caption{
Vertical scale height as a function of the gas vertical velocity dispersion. The solid lines
are computed for the "star-dominated disk" limiting form of the hydrostatic equilibrium
in the disk plane 
(Eq.~\ref{Eq:shei_nonSG}) plotted for a range of possible values 
of the stellar disk parametrization $H_{\star} = 0.3 - 3$~kpc and $\Sigma
_{\star} = 320 - 950 \,$M$_{\odot}$~pc$^{-2}$. The
dashed lines plot the scale height values given by the balance
between the turbulent heating rate and the observed [C{\sc ii}]+H$_2$
line cooling rate (Eq.~\ref{eq:sca_hei_turb_diss}), for two values of the efficiency 
parameter $\epsilon$ (50\% and 100\%, see Sect.~\ref{subsec:scale_height}
for details).
The grey area shows the plausible ($\sigma _g$, $H_g$) parameter space for the 3C~326N gas disk. 
}
\label{fig:scale_height}
\end{figure}

We first consider the case where the stellar gravity dominates over the gas gravity in the disk, 
i.e. $\rho _{\star} \gg \rho _g$. This is likely to be the case for 3C~326N since the stellar mass 
($M _{\star} \approx 3 \times 10^{11}$~M$_{\odot}$) is much higher than the total gas 
mass ($M_g \approx 2 \times 10^{9}$~M$_{\odot}$). In this \textit{"star-dominated disk"}, 
the gravitational force per unit mass of gas along $z$ is constant. Then, 
 the gas scale height is proportional to the velocity dispersion of the gas:
\begin{equation}
\label{Eq:shei_nonSG}
H_g = \frac{1}{2} \left(\frac{H_{\star}}{G \Sigma _{\star} } \right)^{1/2} \sigma _{g} \ ,
\end{equation}
where $H_{\star}$ is the stellar vertical scale height. This expression is plotted in Figure~\ref{fig:scale_height} as the solid 
lines. The parametrization of the stellar morphology, based on our examination of 
SDSS images\footnote{The SDSS images show that 3C~326N has a lenticular 
shape, and the axis ratio of the disk of old stars is similar to the H$_2$ disk seen with SINFONI, 
 although with 10 times the radii (15~kpc).}, is uncertain, 
since we do not know the flattening of the stellar distribution in 3C~326N. 
Therefore, we plot equation~\ref{Eq:shei_nonSG} for four different parametrizations that 
bracket the possible range of stellar scale heights, $H_{\star} = 0.3 - 3$~kpc, and 
stellar mass surface densities, $\Sigma _{\star} = 320 - 950$~M$_{\odot}$~pc$^{-2}$, 
which corresponds to stellar masses 
$M _{\star} = 1 - 3 \times 10^{11}$~M$_{\odot}$ in a sphere of diameter 20~kpc .

Figure~\ref{fig:scale_height} displays the expression of
the scale height constrained by the balance between the turbulent heating rate and the 
observed line cooling rate, given in Eq.~\ref{eq:sca_hei_turb_diss}. 
Eq.~\ref{eq:sca_hei_turb_diss} is plotted as the dashed lines for two values of the 
$\epsilon$ parameter ($\epsilon = 1$ means that 100\% of the turbulent energy 
dissipated is radiated into the [C{\sc ii}] and H$_2$ lines). 
The grey area in Figure~\ref{fig:scale_height} shows the plausible parameter space if 
the gas is within the stellar disk: the gas velocity dispersion would lie 
within the range $60 < \sigma _g < 190$~km~s$^{-1}$, 
and the gas scale height would be comprised within $0.3 < H_g < 4$~kpc. 
This is a plausible solution if 3C~326N is a lenticular galaxy.
We note that gas scale height above $H_{\star} = 3$~kpc are unphysical since in that case  the stars dominate the gravitational potential in the disk.

Then, we consider the case of a \textit{self-gravitating gas disk} in a stellar spheroid morphology, 
where the gravity is dominated by the gas on the scale of the vertical height of the disk ($\rho _g
\gg \rho _{\star}$). Equating $P_{grav} = P_{turb}$, we find that the gas scale 
height can be simplified as:
\begin{equation}
\label{Eq:shei_SG}
H_g = \frac{\sigma _{g}^2}{ \pi G  \Sigma _{g}} \ .
\end{equation}
For the sake of clarity, and because this limiting form is unlikely for 3C~326N ($M_{\star} \gg M_g$, see above), we do not plot this case in Figure~\ref{fig:scale_height}. For the observed value of gas mass surface density in 3C~326N ($\Sigma _{g} = 100\,$M$_{\odot}$~pc$^{-2}$), the self-gravitating disk curve  would intercept the turbulent dissipation 
solution (dashed lines) at gas scale heights much larger than the size of the galaxy ($500 - 900$~kpc for 
$\epsilon = 0.5 - 1$ and $L_{\rm [CII]+H_2} / M_{\rm H_2} = 0.18$~L$_{\odot}$~M$_{\odot}^{-1}$). 
This shows that we cannot reconcile the observed turbulent heating rate with the hypothesis of a self-gravitating disk.

We also consider the case of an elliptical galaxy where the H$_2$ gas is bound by 
the gravitational binding energy of the stellar spheroid
$\frac{3}{5}\frac{G M_{\star}^2}{r}$, where $r$ is the spheroid
radius. We find that the equivalent external pressure on the gas is 
$P_{\rm ext}/k = \frac{3}{10 \pi k} \frac{G M_{\star}^2}{r^4}$.  For a  radius of
10~kpc, $P_{\rm ext}/k = 2 \times 10^6$~K~cm$^{-3}$. This pressure is
comparable to the turbulent pressure $\Sigma _{g} \sigma _{g}^2 /
2 H_g$ observed on the scale of the disk ($P_{\rm ext}/k = 0.8 \times 10^6$~K~cm$^{-3}$
for $H_g=3$~kpc and $\sigma _{g} = 100$~km~s$^{-1}$), and an order of
magnitude higher than the average thermal pressure of the warm H$_2$
gas deduced from the [C{\sc ii}] observations ($10^5$~K~cm$^{-3}$, see
Sect.~\ref{subsec:physicalconditions}). 
Therefore, the corresponding expression of the gas scale height, 
not plotted in Figure~\ref{fig:scale_height}, would intercept the dashed curves of the turbulent 
dissipation solution at locations very similar to the lenticular ("star-dominated disk") solution (grey area).

In conclusion, finding a dynamical state of the gas compatible 
with our interpretation where turbulent heating is the main heating 
source of the molecular gas in 3C~326N and where the H$_2$ and 
[C{\sc ii}] lines are powered by the dissipation of the gas turbulent kinetic 
energy, implies large vertical gas velocity dispersions 
($60 < \sigma _g < 190$~km~s$^{-1}$) 
and high values for the gas scale height ($0.3 < H_g < 4$~kpc). 
It is plausible to find a such high vertical gas velocity dispersion in 3C~326N given the 
observed turbulent velocity dispersion estimated in Sect.~\ref{subsec:turb_dissipation} ($120  < \sigma _{turb} < 330$~km~s$^{-1}$).
From the near-infrared H$_2$ line morphology, the H$_2$ disk has to have 
a thickness $H_g < 0.9$~kpc \citep{Nesvadba2011a}. However, the 
near-IR ro-vibrational H$_2$ lines trace denser and warmer gas than the 
[C{\sc ii}] line, accounting for a small fraction of the total H$_2$ mass, 
so the [C{\sc ii}]-emitting disk may be larger. The SINFONI 
observations also have a limited surface brightness sensitivity, so we 
may not see all the warm H$_2$ disk.

\section{Discussion: turbulence as a feedback mechanism}
\label{sec:feedback}

In this section we discuss some general implications of the work 
presented in this paper for galaxy evolution.

The [C{\sc ii}] profile we observe in 3C~326N exhibits very broad wings,
especially toward the blue where the velocities reach up
to about $-$1000~km~s$^{-1}$. Such high velocities are typically
interpreted as terminal velocities of energetic outflows. Because
of the relatively high densities necessary for observing strong [C{\sc
ii}] emission \citep[e.g.,][]{Maiolino2012}, CO emission and absorption
\citep[e.g.,][]{Feruglio2010, Dasyra2012, Cicone2012, Morganti2013a, Cicone2014}, and molecular
absorption lines \citep[e.g.,][]{Aalto2012, Spoon2013}, large mass ejection 
rates have usually been derived from these outflow velocities. 

These high gas velocities are consistent with the velocities modeled in jet
interactions with gas disks surrounding an AGN \citep{Gaibler2012,
Wagner2012, Wagner2013}.  In these simulations, the high gas velocities
are, however, associated with gas at lower densities than we observe in the
[C{\sc ii}] emitting gas \citep{Gaibler2012}.  Getting agreement between
the velocities of the different gas phases requires the warm molecular gas
to be formed in the turbulent backflow of the jet, and its formation 
of molecular gas could be triggered by the compression
generated by the two-sided radio jet. If so, this implies that not only
does the jet play a crucial role in energizing the turbulent cascade, but
in fact it is also responsible for the physical state (predominately molecular
instead of hot ionized or warm neutral medium) and the distribution of the
molecular gas (its disk-like structure). Despite the
high velocities observed in the wings of the line profiles of [C{\sc
ii}] and the ro-vibrational lines of H$_2$, the molecular gas is perhaps confined by the
pressure of the hot halo gas with which the jet is strongly interacting.

A natural outcome of this picture is that the gaseous disks in
radio galaxies should be approximately perpendicular to the jet axis
\citep{Dekoff1996, Martel1999, Dekoff2000, Carilli2000}, although many sources
would show more complex gas distributions depending on the phase probed
or the symmetries in the jets.  Moreover, if the jets of the AGN shut
down, the phase and structure of the disk will change dramatically,
perhaps even experiencing a phase transition from a molecular phase to a hot ionized phase.

In addition, the picture of a highly turbulent molecular disk may also
explain one of the most significant puzzles in relation to outflows from
AGN, namely, the apparent short gas ``erosion'' timescales in some of
the sources, less than $\approx$10 Myrs.  A large percentage of the AGN population show
such short timescales \citep[e.g.,][]{Feruglio2010, Nesvadba2011a, Cicone2014}.
In fact, \citet{Nesvadba2011a} find that the erosion timescale of the
3C~326N disk was about that of the orbital timescale, which roughly
agrees with our estimate ($3-10 \times 10^7$~yrs). These findings beg the question: why
do we see outflows and gas disks at all in these sources?  Either the
outflow duty cycle is short and the gas  replenished quickly, or the
erosion timescales are wrong \citep[for example, the total mass estimated
or the underlying model for estimating the outflow rate;][]{Alatalo2011, Alatalo2014b}.

As we discussed earlier, the interaction between the backflow, generated
by the propagating jets, and the disk can create high turbulent velocities,
in the range $100-1000$~km~s$^{-1}$ \citep{Gaibler2012}.  
Because the broad line wings have several times the velocity dispersion of the majority
of the gas, they could be the signature of the stochastic nature of turbulence 
\citep[see, for example,][]{Falgarone1990, Pety2003}.
Since the pressure of the overlying hot gas is probably several times
that of the disk itself \citep{Gaibler2012}, this high velocity gas is
confined and not outflowing because it is generated by the interaction with
the hot gas phase and can be naturally related to the turbulent cascade, 
depending on the exact source of the high velocity.  The
high velocities do not necessarily indicate escaping gas and may well be a
natural outcome of our overall picture of a high turbulent molecular disk
generated by the action of the radio jet on the host ISM and halo gas.

One of the other major implications of these results is that the
energy injection from the radio source into the ISM can be both
``positive'' and ``negative'' feedback. The feedback is positive
in the sense that the over-pressure generated by the jets allows for
enhanced formation of H$_2$ gas and the creation of large molecular
disks around AGN \citep{Wolfire1995}.  However, radio loud AGN may maintain
the molecular gas in such a state that it cannot form stars efficiently
owing to the high turbulence.  Thus we
caution against any interpretation of enhanced molecular gas formation around
AGN as being systematically a sign of positive feedback in the sense of enhanced star formation.

It appears from our analysis of 3C~326N that while the high compression
of the disk resulting from the backflow and turbulence generated
by two-sided jets may lead to strong cooling, i.e., enhanced [C{\sc ii}]
emission, and a significant fraction of the gas is molecular, it may
also suppress star formation by only allowing a small fraction of the
gas to become self-gravitating on any scale. This is probably due to the exceptional 
turbulent support in 3C~326N, which maintains a very large disk thickness, and
  is able to stabilize the disk on large spatial scales against fragmentation into bound star-forming
clumps of gas. This offers a natural explanation for the quenching of
star formation in 3C~326N.  
The compression may be sufficient to allow small gas clumps to become self-gravitating
and collapse to form stars, but not at the rate expected given the
gas surface densities observed.  We are only suggesting that the star
formation efficiency is much lower when the gas phase mixture and physical structure
of molecular disks are related by the energy input from radio jets.

This is, however, not to say that in all cases, the action of the jet
on a gaseous disk always inhibits star formation.  If the disk were
pre-existing, rather than being formed by the action of the jet itself as we have
proposed, and were composed of gas and stars distributed in a disk,
the outcome of the interaction with the backflow might well have been different
\citep[e.g.,][]{Martig2013}.  In disk galaxies, because the stars dominate
the baryonic mass distribution, the stellar disk is the main source of
(self-)gravity.  In such a case, the gas will be initially gravitationally
bound and the jet over-pressure may enhance star formation.  We suggest
this picture to emphasize the role played by self-gravity on large scales.
In fact, as we have found, the action of the jet on the surrounding medium
is to increase the overall pressure and make the conditions such that 
high molecular fractions are likely \citep{Wolfire1995}. This may then lead
to an enhanced star formation efficiency in disk structure that is already
self-gravitating \citep[][]{Silk2009, Silk2013}.

\section{Summary of results}
\label{sec:summary}

We detected  [C{\sc ii}]$\lambda \, 158\mu$m line emission from the 
3C~326N radio galaxy with the \textit{Herschel/PACS} instrument. The [C{\sc ii}] is 
very luminous, twice as bright as the H$_2$ 0-0S(1) line detected by the 
\textit{Spitzer IRS}. 
The [C{\sc ii}]-to-FIR and [C{\sc ii}] to PAH$\,7.7\mu$m flux ratios are 
extremely high ($0.035 \pm 0.014$ and $>2.5$), 3 to 30 times 
higher than what is observed in typical star-forming galaxies, showing that UV 
is not the main excitation process.

The [C{\sc ii}] line profile is broad and asymmetric, with a core of 
intrinsic $\mbox{FWHM} = 340$~km~s$^{-1}$, and a pronounced blueshifted wing of 
$\mbox{FWHM} = 800$~km~s$^{-1}$. The line is very likely shaped by a mixture of systematic 
rotation, outflowing gas, and turbulence. We compared the [C{\sc ii}] kinematics  
to other tracers to discuss these three relative contributions to the line 
profile. The [C{\sc ii}] line matches the near-IR H$_2$,
H$\alpha$, Na~D absorption, and CO line profiles well, although these
lines probe different phases and mass fractions of the gas. 
 Attributing the line wing
 to a wind leads to high outflow rates in the range of
$20-50$~M$_{\odot}$~yr$^{-1}$, depleting the gas reservoir in timescales
of $30-100$~Myr, within an order of magnitude of the orbital
timescale. If true, we are observing this object at a very specific and brief 
time of its evolution, if the disk is not replenished. 
These outflow rates may be overestimated because turbulence (generated by the 
interaction between the disk and the jet's backflow) can produce 
short-lived, large velocity increments that contribute to the skewness 
of the line profile and could mimic outflowing gas.

By combining infrared \textit{Spitzer} and \textit{Herschel} photometry and 
spectroscopy with gas and dust modeling, we inferred the physical and chemical 
conditions in the [C{\sc ii}]-emitting gas. We concluded that the [C{\sc ii}] line emission, 
like the infrared rotational H$_2$ lines, traces a significant fraction (30 to 50\%) 
of the total mass of molecular gas, which is warm ($70 <T < 100$~K), and where most of the 
atomic Carbon is singly ionized. The high observed [C{\sc ii}] to [O{\sc i}] ratio 
allowed us to constrain the average molecular gas density within the 
range $700 < n_{\rm H} < 3000$~cm$^{-3}$. 

We showed that the most plausible heating source of the molecular gas is the
dissipation of supersonic turbulence.  This turbulence must be sustained
by a vigorous energy source because the radiative cooling time of the
warm molecular gas is short ($\approx 10^4$~yr).  Maintaining this turbulent heating 
only requires a small percentage of the jet mechanical energy, and the time to dissipate 
this kinetic energy of the warm molecular gas is about several 10~Myr up to about 100~Myr (Eq.~\ref{eq:turb_diss_time}). 
Cosmic ray heating would require warm H$_2$ densities above $3000$~cm$^{-3}$ 
for the gas to remain molecular in the presence of cosmic ray ionization. Cosmic rays 
are therefore not the favored source of heating, but they are very likely to play
an important chemical role in maintaining a high C$^+$/CO fractional
abundance in the molecular gas. This contrasts with what is
normally observed  in star-forming galaxies where the gas is heated by
photoelectric heating of the cold neutral medium.

We estimated the dynamical state of the gas disk that is compatible 
with our interpretation where turbulent dissipation is the main heating 
source of the molecular gas in 3C~326N, as traced by H$_2$ and 
[C{\sc ii}] lines. We found that, owing to the large gas vertical velocity dispersions required to sustain the 
turbulent heating rate ($60 < \sigma _g < 190$~km~s$^{-1}$), the exceptional turbulent 
support maintains high values of the gas scale height ($0.3 < H_g < 4$~kpc). 
Therefore, radio-loud AGN can enhance the formation of molecular gas ("positive" feedback) 
because of the over-pressure generated by the jet, but in extremely turbulent environments like 
3C~326N, they can also prevent the disk 
from becoming self-gravitating on any scale ("negative" feedback). 
This offers a natural explanation for the quenching of
star formation in 3C~326N. Indeed, 
3C~326N has a substantial amount of molecular gas 
($2 \times 10^9$~M$_{\odot}$, inferred from CO(1-0) line observations and 
modeling of the far-infrared dust emission), despite a 
very low star formation rate (SFR$<0.07$~M$_{\odot}$~yr$^{-1}$). 
Strong turbulent heating appears to be an
efficient negative feedback mechanism, and a significant removal of cold
gas by the AGN is not necessarily needed to prevent the gravitational
collapse of dense gas. The star formation efficiency can be much lower when the 
phase and physical structure of molecular disks are affected by the mechanical 
energy input from radio jets.

\begin{acknowledgements}

We wish to thank Nicole Nesvadba for proposing these observations and in
generously sharing hers with us. We also thank Volker Gaibler
for discussing the results and nature of his radio jet simulations and Heino
Falcke and Peter Biermann for their insights into cosmic rays in the
environments of AGN. 
We are also very grateful to the NHSC staff for providing remote computing 
sources and help with the PACS data processing.  
PACS has been developed by a consortium of institutes led by MPE (Germany)
 and including UVIE (Austria); KU Leuven, CSL, IMEC (Belgium); CEA, LAM (France); 
 MPIA (Germany); INAF-IFSI/OAA/OAP/OAT, LENS, SISSA (Italy); IAC (Spain). 
 This development has been supported by the funding 
 agencies BMVIT (Austria), ESA-PRODEX (Belgium), CEA/CNES (France), DLR 
 (Germany), ASI/INAF (Italy), and CICYT/MCYT (Spain).
The Open University is incorporated by Royal Charter (RC 000391), an exempt 
charity in England \& Wales and a charity registered in Scotland (SC 038302).
\end{acknowledgements}

\bibliographystyle{aa.bst}
\bibliography{3C326_CII.bbl}

\end{document}